\documentclass{article}

\PassOptionsToPackage{numbers, compress}{natbib}
\usepackage[eandd, final]{neurips_2026}

\usepackage[utf8]{inputenc} % allow utf-8 input
\usepackage[T1]{fontenc}    % use 8-bit T1 fonts
\usepackage{hyperref}       % hyperlinks
\usepackage{url}            % simple URL typesetting
\usepackage{booktabs}       % professional-quality tables
\usepackage{amsfonts}       % blackboard math symbols
\usepackage{nicefrac}       % compact symbols for 1/2, etc.
\usepackage{microtype}      % microtypography
\usepackage{xcolor}         % colors

\usepackage{multirow}
\usepackage{graphics}

\usepackage{colortbl}
\usepackage{graphicx}
\usepackage[table]{xcolor}
\usepackage{diagbox}
\usepackage{tabularx}              % 자동 열 너비 조절을 위해
\usepackage{booktabs}              % 깔끔한 테이블 선을 위해
\usepackage{array}                 % 열 정렬을 위해
\usepackage{kotex}[cjk]
\usepackage{subcaption}
\usepackage{geometry}
\usepackage{tabularx} % 폭 자동 조절을 위해 필요
\usepackage{amsmath}
\usepackage{wrapfig}
\usepackage{arydshln}

\title{\textsc{MLAIRE}: Multilingual Language-Aware \\ Information Retrieval Evaluation Protocol}

\author{Youngjoon Jang$^{1}$, Seongtae Hong$^{1}$, Hyeonseok Moon$^{2}$\thanks{Corresponding authors.}, Heuiseok Lim$^{1}$\footnotemark[1] \\
$^{1}$Department of Computer Science and Engineering, Korea University \\
$^{2}$Sookmyung Women's University \\
\texttt{\{dew1701, ghdchlwls123, limhseok\}@korea.ac.kr} \\
\texttt{hyns.moon@sookmyung.ac.kr}
}

\begin{document}

\maketitle

\begin{abstract}
Multilingual Information Retrieval is increasingly important in real-world search settings, where users issue queries over mixed-language corpora. Existing evaluations mainly reward language-agnostic semantic relevance, treating relevant passages equally regardless of language. Yet retrieval utility also depends on the language of the retrieved passages: users may prefer results they can read and verify in the query language, and query--passage language mismatch can complicate downstream grounding and answer verification in Retrieval-Augmented Generation systems. To evaluate this language-aware dimension, we introduce \textbf{\textsc{MLAIRE}}, a \textbf{M}ultilingual \textbf{L}anguage-\textbf{A}ware \textbf{I}nformation \textbf{R}etrieval \textbf{E}valuation protocol that disentangles cross-lingual semantic retrieval from query-language preference. \textsc{MLAIRE} constructs controlled pools with parallel passages across languages, enabling measurement of semantic retrieval accuracy and query-language preference when equivalent translations are available.
We propose language-aware metrics, including Language Preference Rate (LPR) and Lang-nDCG, together with a 4-way decomposition separating semantic and query-language preference failures. Evaluating 33 dense, sparse, and late-interaction retrievers, we show that standard metrics obscure distinct behaviors: semantically strong retrievers may return correct content in a non-query language, while retrievers with stronger query-language preference may retrieve less semantically relevant passages.
Our datasets\footnote{\url{https://huggingface.co/collections/yjoonjang/mlaire}} and code\footnote{\url{https://github.com/yjoonjang/MLAIRE}} are publicly available.
\end{abstract}

\section{Introduction}

Multilingual Information Retrieval (MLIR) aims to retrieve relevant content when queries and documents appear in diverse languages~\citep{nie2002multilingual, peters2012multilingual}. 
This setting is central to real-world search and Retrieval-Augmented Generation (RAG) systems, where users issue queries over corpora containing content in many languages. 
Accordingly, MLIR systems are primarily evaluated by whether they can identify semantically relevant content across languages~\citep{Yang_2024, yang2024language, yang2025language}.

However, semantic retrieval quality alone does not determine whether a retrieved result is useful to the user. 
When several language versions of the same relevant content are available, a retriever may rank non-query-language content above its query-language counterpart.
This creates a gap between semantic retrieval and query-language preference: the former asks whether the retrieved content is relevant, while the latter asks whether the relevant content is written in the query language.

\begin{wrapfigure}[16]{r}{0.5\textwidth}
    \vspace{-0.16cm}
    \centering
    \includegraphics[width=\linewidth]{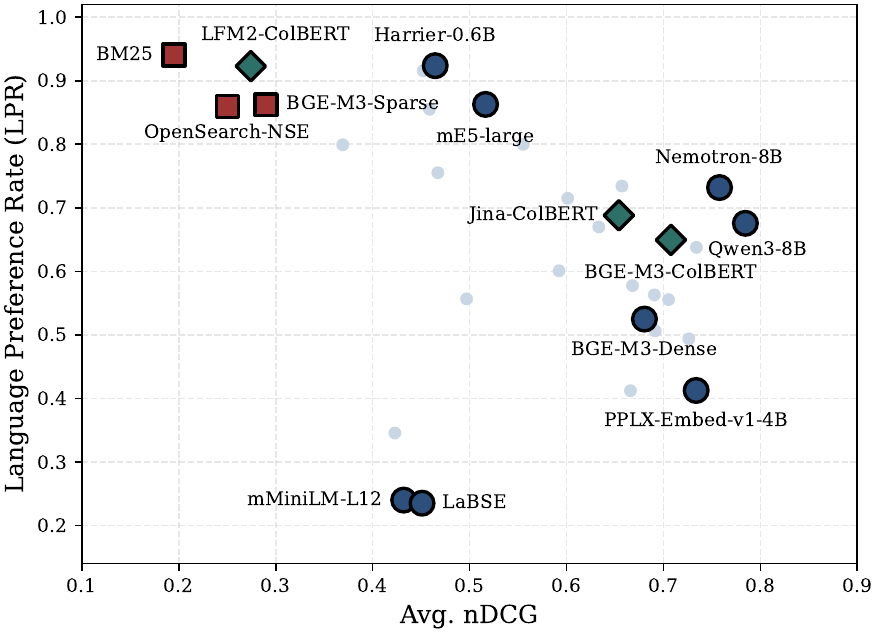}
    \caption{
    Each point shows a retriever evaluated on \textsc{MLAIRE}, with average nDCG on the x-axis and Language Preference Rate (LPR) on the y-axis.
    }
    \label{fig:ndcg_lpr_mismatch}
    \vspace{-0.45cm}
\end{wrapfigure}

This gap is visible even among strong multilingual retrievers. 
The scatter plot in Figure~\ref{fig:ndcg_lpr_mismatch} compares standard nDCG with Language Preference Rate (LPR), which measures whether the query-language version of the target content is scored above its semantically equivalent alternatives. The results are macro-averaged over the three \textsc{MLAIRE} datasets.
In this figure, PPLX-Embed-v1-4B and BGE-M3-Dense achieve strong standard retrieval performance but show lower LPR than mE5-large, while BM25 and OpenSearch-NSE exhibit high LPR despite weaker nDCG. 
This suggests that standard retrieval metrics can obscure whether a retriever prioritizes query-language evidence.

\setlength{\columnsep}{8pt}
\setlength{\intextsep}{4pt}
This distinction matters in practical multilingual retrieval scenarios. 
The language of the retrieved passage affects whether users can read, verify, and act on the result. 
It can also affect downstream RAG behavior: when the query and retrieved passages are written in different languages, the generator must interpret cross-lingual evidence while producing an answer in the user's language. 
Recent studies on multilingual and cross-lingual RAG report that query--context language mismatch can degrade answer correctness and make models less likely to preserve the expected response language~\citep{park2025investigating, liu2025xragcrosslingualretrievalaugmentedgeneration, ki2025linguistic}. 
Thus, query-language preference in retrieval is not merely a user-facing preference, but also a retrieval behavior that can affect downstream answer generation.

To evaluate this behavior, we introduce \textbf{\textsc{MLAIRE}}, a Multilingual Language-Aware Information Retrieval Evaluation protocol.
Inspired by the parallel-corpus formulation of~\citet{roy2020lareqa}, \textsc{MLAIRE} constructs candidate pools where semantically equivalent passages coexist across languages. 
Whereas LAReQA~\citep{roy2020lareqa} treats all translations of the relevant passage as equally relevant, \textsc{MLAIRE} distinguishes the query-language counterpart from its semantically equivalent translations. 
This setup preserves standard MLIR evaluation of cross-lingual semantic retrieval while making query-language preference directly observable. 
Using \textsc{MLAIRE}, we report conventional retrieval metrics together with language-aware diagnostics, including LPR, Lang-nDCG, and a 4-way decomposition of rank-1 outcomes.
Our analysis shows that query-language preference constitutes a distinct and structured dimension of MLIR behavior, rather than a property captured by language-agnostic retrieval metrics alone.

Our main contributions are as follows:
\begin{itemize}
    \item We formulate query-language preference as an important aspect of MLIR evaluation.
    \item We introduce \textbf{\textsc{MLAIRE}}, a language-aware evaluation protocol with metrics and diagnostics for analyzing retrieval behavior beyond semantic relevance.
    \item We evaluate 33 retrievers across dense, sparse, and late-interaction architectures, revealing systematic mismatches between semantic retrieval quality and query-language preference.
\end{itemize}

\section{Background}
\label{sec:background}
\subsection{Multilingual Information Retrieval (MLIR)}
\paragraph{Definition of MLIR}
Information Retrieval (IR) settings can be distinguished by the language relationship between the query and the candidate corpus~\citep{braschler2002clef, lawrie2025neuclirbench}. 
Monolingual IR assumes that queries and relevant documents are written in the same language~\citep{thakur2021beirheterogenousbenchmarkzeroshot, muennighoff2023mteb}, while Cross-Lingual IR (CLIR) considers settings where the query and target documents are written in different languages~\citep{vulic2015monolingual, lee2026clear}. 
In this paper, we adopt the definition of MLIR following the Cross-Language Evaluation Forum (CLEF): a task in which queries are issued in different languages and the candidate pool may contain relevant passages in multiple languages simultaneously~\citep{braschler2002clef}. 
Unlike Multi-Monolingual IR, where each query is evaluated against a same-language corpus~\citep{enevoldsen2025mmteb}, MLIR allows passages in multiple languages to coexist within a shared candidate pool.

\paragraph{Conventional Evaluation of MLIR}
Existing MLIR evaluations primarily adopt a language-agnostic view of relevance. 
A retrieved passage is considered relevant if it satisfies the information need, regardless of the language in which it is written~\citep{roy2020lareqa, lawrie2025neuclirbench, hong2026improving}. 
This view is essential for evaluating cross-lingual semantic retrieval: a retriever should be able to recognize that semantically equivalent texts in different languages express the same information. 
This objective is consistent with multilingual representation learning, which aims to place semantically equivalent texts from different languages in a shared embedding space~\citep{chi2021infoxlm, feng2022language}.

However, this language-agnostic view does not distinguish between different language versions of the same relevant content.
When the query-language passage and its non-query-language translations are all semantically relevant, standard metrics such as Recall and nDCG treat them as equally relevant~\citep{buckland1994relationship, valizadegan2009learning}.
As a result, existing evaluations can determine whether a retriever finds the right content, but not whether it prioritizes the version written in the query language.
We therefore evaluate query-language preference as a complementary axis alongside language-agnostic semantic retrieval.

\subsection{Query-Language Preference as an Evaluation Dimension}
Query-language preference matters because the language of a retrieved passage is part of retrieval utility.
A passage can be semantically relevant but still difficult for the user to read, verify, or act on if it is written in a language the user did not use or cannot readily understand.
This issue is especially important in user-facing search, where prior studies on multilingual production search systems show that users prefer results written in the language of their query~\citep{csa2020, singh2007language, steichen2018, steichen2021, google2023multilingual}.
% ~\citep{csa2020cant, singh2007query, steichen2021multilingual, ling2018query, nahar2023search}.
In this sense, query-language preference reflects a legitimate aspect of user intent.

The same issue also arises in Retrieval-Augmented Generation (RAG) systems, where retrieved passages are directly consumed by a generator.
If the query and retrieved passage are written in different languages, the generator must interpret cross-lingual evidence while producing an answer in the user's language.
Recent studies on multilingual and cross-lingual RAG report that query--context language mismatch can degrade answer correctness and make models less likely to preserve the expected response language~\citep{park2025investigating, liu2025xragcrosslingualretrievalaugmentedgeneration, ki2025linguistic}.

\begin{figure}[t]
    \centering
    \includegraphics[width=\textwidth]{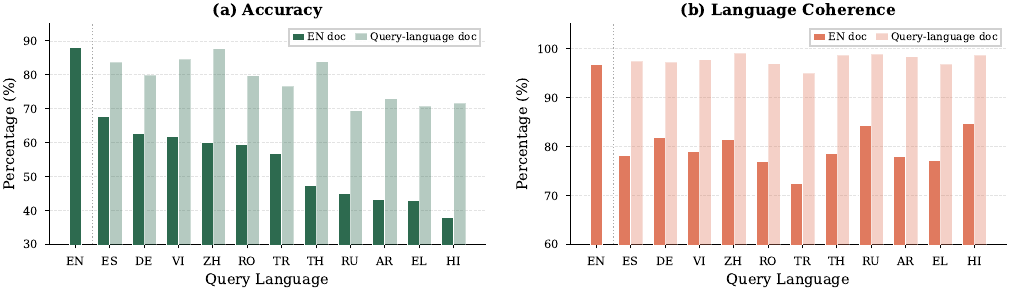}
    \caption{
    RAG experiment on XQuAD across 12 query languages. 
    We evaluate Qwen2.5-7B-Instruct using Exact Match~(EM) and response-language correctness.
    \textit{EN doc} provides the gold passage in English, and \textit{Query-language doc} provides the same gold passage written in the query language.
    }
    \label{fig:rag_language_mismatch}
    \vspace{-0.4cm}
\end{figure}

To further examine this effect, we conduct a controlled RAG experiment using XQuAD~\citep{artetxe2020cross} and Qwen2.5-7B-Instruct~\citep{qwen2025qwen25technicalreport}.
For each query language, we compare two conditions with gold passages of the same meaning: an English passage and a passage written in the query language.
Figure~\ref{fig:rag_language_mismatch} reports answer accuracy in (a) and language coherence in (b), where language coherence indicates whether the model answers in the query language.
When the query is in English, providing the English relevant passage yields high accuracy and high language coherence.
However, when non-English queries are paired with English relevant passages of the same meaning, both answer accuracy and language coherence drop substantially.
Replacing the English passage with the query-language passage consistently improves performance and makes the model much more likely to answer in the query language.
Since the two conditions provide equivalent gold content, this gap shows that the language of the retrieved passage matters even when semantic relevance is controlled.

These observations motivate a language-aware evaluation perspective that measures whether retrievers prioritize query-language evidence when equivalent relevant passages are available across languages.

\section{\textsc{MLAIRE}}
\label{sec:mlaire}

\textsc{MLAIRE} follows a simple design principle: each query is paired with a candidate pool that contains semantically equivalent relevant passages in multiple target languages.
This construction creates a controlled retrieval setting where a model's ranking behavior reveals two complementary properties.
The first is its ability to retrieve semantically relevant content across languages, and the second is its tendency to prioritize query-language evidence when multiple relevant translations are available.

\subsection{Evaluation Dataset Construction}
\label{sec:data}

We build the evaluation dataset from three multilingual resources containing (partially) parallel passages.
For each dataset, we organize passages into content groups, where each group contains passages expressing the same underlying content in different languages. 
For a query \(q\), all passages in its target content group are treated as semantically relevant, regardless of language. 
Among them, the passage written in the same language as \(q\) is treated as the query-language relevant passage. 

\begin{wraptable}{r}{0.42\textwidth}
\vspace{-0.2cm} % pull the table up to align with the first line of text
\centering
\caption{Summary of the evaluation datasets used in \textsc{MLAIRE}}
\label{tab:benchmark-stats}
\footnotesize
\setlength{\tabcolsep}{4pt}
\begin{tabular}{l rrr}
\toprule
 & Belebele & XQuAD & MLQA \\
\midrule
\# Languages     & 122       & 12       & 7        \\
\# Queries       & 109{,}800 & 14{,}280 & 42{,}245 \\
\# Passages      & 59{,}536  & 2{,}880  & 35{,}145 \\
Max $N_\text{rel}$ & 122     & 12       & 4        \\
Parallelism      & Full      & Full     & Partial  \\
\bottomrule
\end{tabular}
% \vspace{-1em} % tighten the gap between the table and the wrapped text below
\end{wraptable}

\textbf{Belebele}~\citep{bandarkar2024belebele} provides 488 reading-comprehension passages derived from FLORES-200~\citep{costa2022no}, each translated into 122 language variants, with associated queries.
\textbf{XQuAD}~\citep{artetxe2020cross} provides 240 parallel paragraphs and 1,190 extractive QA pairs in each of 12 languages.
\textbf{MLQA}~\citep{lewis2020mlqa} provides partially parallel extractive QA data across 7 languages; unlike Belebele and XQuAD, its passages are not fully parallel, and most groups contain translations in only 3--4 languages rather than all 7. We note that Belebele, XQuAD, MLQA are all utilized in MMTEB~\citep{enevoldsen2025mmteb}. Table~\ref{tab:benchmark-stats} summarizes the evaluation dataset construction of \textsc{MLAIRE}.

% ---- 3.2 Language-Aware Metrics ----
\subsection{Language-Aware Metrics}
\label{sec:metrics}

We report standard retrieval metric (nDCG@$k$) and language-aware metrics designed to capture query-language preference behavior.
For a query \(q\), let \(\ell_q\) denote its language and \(g_q\) denote its target content group.
Each content group consists of semantically equivalent passages across languages.
For a passage \(d\), let \(\ell_d\) and \(g_d\) denote its language and content group, respectively.

\paragraph{Language Preference Rate (LPR)}
Let \(Q\) denote the set of evaluation queries.
For each query \(q \in Q\), let \(d_q^{*}\) denote the highest-scoring passage among passages in the target content group:
\[
d_q^{*}
=
\operatorname*{arg\,max}_{d: g_d = g_q} s(q,d),
\]
where \(s(q,d)\) is the retriever score.
We define LPR as
\begin{equation*}
\mathrm{LPR}
=
\frac{1}{|Q|}
\sum_{q \in Q}
\mathbf{1}\!\left[\ell_{d_q^{*}} = \ell_q\right]
\end{equation*}
LPR measures how often the query-language version of the target content is scored above its cross-lingual alternatives.

\paragraph{Lang-nDCG@$k$}
To evaluate language-aware ranking quality, we assign higher relevance to passages that both match the target content group and are written in the query language:
\noindent
\begin{minipage}{0.49\linewidth}
\begin{equation*}
\mathrm{rel}_{\mathrm{lang}}(d,q)
=
\begin{cases}
3, & \text{if } g_d = g_q \text{ and } \ell_d = \ell_q,\\
2, & \text{if } g_d = g_q \text{ and } \ell_d \neq \ell_q,\\
0, & \text{otherwise}
\end{cases}
\end{equation*}
\end{minipage}
\hfill
\begin{minipage}{0.49\linewidth}
\begin{equation*}
\mathrm{DCG}@k
=
\sum_{i=1}^{k}
\frac{2^{\mathrm{rel}_{\mathrm{lang}}(d_i, q)} - 1}{\log_2(i + 1)}
\end{equation*}
\end{minipage}

\vspace{0.5em}
We compute Lang-nDCG@$k$ by normalizing DCG@$k$ with the maximum possible DCG (IDCG) under this language-aware grading scheme.
Unlike standard nDCG@$k$, Lang-nDCG@$k$ distinguishes the query-language version of the target content from its cross-lingual alternatives.

% \subsection{4-way failure decomposition}
% To diagnose retrieval behavior, we classify each retrieved passage
% along two binary axes: (i) whether the passage belongs to the correct semantic group, and (ii) whether it matches the query language. This yields four mutually exclusive outcomes:
% (1) \texttt{perfect} if semantically relevant and in the same language;
% (2) \texttt{lang\_fail} if semantically relevant but in a different language;
% (3) \texttt{sem\_fail} if in the same language but semantically irrelevant;
% (4) \texttt{both\_fail} if neither semantically relevant nor in the same language. For a given cutoff $k$, we apply this classification to every passage in the top-$k$ retrieval and aggregate proportions across all queries; with $k{=}1$ this recovers the standard rank-1 error analysis, while
% larger $k$ (e.g., $k{=}20, 200$) reveals the composition of the entire returned list---exposing, for instance, retrievers that fill the top-$k$ with same-language but irrelevant distractors. Unlike standard relevance-based metrics, which treat \texttt{perfect}
% and \texttt{lang\_fail} equally as relevant, this decomposition makes the source of retrieval errors explicit.
\subsection{Top-1 4-way Decomposition}
\label{sec:top1_4way_decomposition}

To diagnose retrieval behavior, we classify the top-1 ranked passage by semantic relevance and query-language match.
This yields four mutually exclusive outcomes:
(1) \texttt{perfect}, if the passage is semantically relevant and in the query language;
(2) \texttt{lang\_fail}, if it is semantically relevant but in a different language;
(3) \texttt{sem\_fail}, if it is in the query language but semantically irrelevant; and
(4) \texttt{both\_fail}, if it is neither semantically relevant nor in the query language.
Aggregating these outcomes across queries reveals whether rank-1 errors are primarily driven by language mismatch, semantic mismatch, or both.
This analysis complements LPR: while LPR compares language preference among semantically equivalent target passages, the rank-1 decomposition characterizes the actual first result returned from the full candidate pool.

\section{Experimental Setup}
\label{sec:experimental-setup}

% ---- 4.1 Retrieval Protocols ----

\subsection{Retrieval Paradigms}
\label{sec:protocols}

We evaluate 33 publicly available multilingual retrievers across three retrieval paradigms: dense, sparse, and late-interaction.
This pool covers a broad suite of multilingual embedding models ranging from 100M to 8B parameters, alongside three sparse and three late-interaction retrievers.
Our selection prioritizes breadth across paradigms and model scales, and includes widely used or recently released publicly available multilingual retrievers at the time of our experiments.

\paragraph{Dense}
Our dense retrievers encompass diverse model lineages, ranging from widely adopted encoder-only families such as multilingual-e5~\citep{wang2024multilingual}, bge-m3~\citep{chen2024bge}, gte~\citep{zhang2024mgte}, snowflake-arctic~\citep{yu2024arctic}, nomic-embed~\citep{nussbaum2024nomic}, embeddinggemma~\citep{vera2025embeddinggemma} and jina~\citep{sturua2024jina, akram2026jina}, to recent LLM-based embedding models including Qwen3-Embedding~\citep{zhang2025qwen3}, llama-nemotron~\citep{moreira2024nv}, and pplx-embed~\citep{eslami2026diffusion}.
In this paradigm, queries and passages are independently encoded into fixed-dimensional vectors and scored by cosine similarity.
We use each model's prescribed pooling strategy (CLS, mean, or last-token) and follow their recommended instruction templates; representative prefix formats are listed in Appendix~\ref{app:instruction-templates}.

\paragraph{Sparse}
We evaluate three multilingual sparse retrievers: a subword lexical baseline and two learned sparse models.
For BM25, we tokenize queries and passages with the XLM-RoBERTa tokenizer~\citep{conneau2020unsupervised} and index with Lucene-style BM25 ($k_1{=}1.2$, $b{=}0.75$)~\citep{bialecki2012apache}.
For the opensearch-neural-sparse-encoding-multilingual-v1~\citep{geng2025competitivesearchrelevanceinferencefree}, queries and passages are encoded into learned sparse vectors over the MLM vocabulary, and scored by inner product.
For bge-m3-sparse, we use the lexical-weight head of bge-m3~\citep{chen2024bge}, which assigns a weight to each input token, and score a query--passage pair by the sum of weight products over their shared tokens.

\paragraph{Late-Interaction}
For late-interaction retrieval, we evaluate jina-colbert-v2~\citep{xiao-etal-2024-jina}, LFM2-ColBERT-350M~\citep{liquidai2025lfm2}, and bge-m3-colbert, the multi-vector head of bge-m3~\citep{chen2024bge}, all of which build upon the late-interaction architecture~\citep{khattab2020colbert}.
Queries and passages are represented as token-level vectors and scored with MaxSim, which sums the maximum similarity between each query token and all passage tokens.
For scalable and efficient evaluation, jina-colbert-v2 and LFM2-ColBERT passages are indexed using the PLAID engine~\citep{santhanam2022plaid}; for bge-m3-colbert, which is not distributed in the format of our PLAID toolchain, we compute exact MaxSim over the full candidate pool.

% We score each query--passage pair by computing the MaxSim: for each query token, we take the maximum cosine similarity to any passage token, then sum across query tokens.

% ---- 4.3 Evaluation Protocol ----
\subsection{Evaluation Protocol}
\label{sec:eval-protocol}

We evaluate each retriever on the three datasets of \textsc{MLAIRE} independently: for every (model, dataset) pair, we encode the dataset's full corpus and retrieve the top-$k$ passages per query.
The retrieval depth $k$ is chosen per dataset so that it exceeds the maximum number of relevant passages per query: we use $k{=}20$ for MLQA (at most 4 relevant passages per query) and XQuAD (12 relevant passages per query), and $k{=}200$ for Belebele, whose 122-language parallel structure admits up to 122 relevant passages per query.
Because LPR compares the relative scores of semantically equivalent target passages, we compute LPR using scores for all passages in the target content group, independent of whether those passages appear in the retrieved top-$k$ list.
Full hardware configuration, dependency partitioning across virtual environments, and reproducibility artifacts are described in Appendix~\ref{app:infrastructure}.

\section{Results and Analysis}
\label{sec:results}

We evaluate 33 retrievers on \textbf{\textsc{MLAIRE}}.
Our analysis shows that query-language preference is an independent behavioral axis associated with semantic alignment, model family and scale, and lexical anchoring, which is the tendency to prefer passages that share surface tokens with the query.
% Section~\ref{sec:main-results} first shows that standard language-agnostic retrieval metrics are weakly aligned with LPR.
% Section~\ref{sec:scoredist} decomposes rank-1 retrieval into semantic and language-level failures, revealing why models with similar aggregate scores can fail in qualitatively different ways.
% Section~\ref{sec:per-lang-lpr} then analyzes per-query-language LPR on XQuAD and shows that language leakage is concentrated in specific language groups.
% Finally, Section~\ref{sec:resource-gravity} extends the analysis to 122 languages in Belebele and examines where wrong-language relevant passages come from when LPR fails.

\begin{table*}[t]
	\centering
	\caption{Retrieval results across three datasets for all 33 retrievers. MLQA (7 languages) and XQuAD (12 languages) are evaluated at $k{=}20$; Belebele (122 languages) is evaluated at $k{=}200$. All values are percentages (raw scores $\times 100$). \textbf{Base} is standard nDCG; \textbf{Lang} is language-aware nDCG (Lang-nDCG). Per column, the best value is \textbf{bold} and the second best is \underline{underlined}.}
	\label{tab:main-results}
	\renewcommand{\arraystretch}{1.3}
	\setlength{\tabcolsep}{10pt}
	\resizebox{\textwidth}{!}{
	\begin{tabular}{l | cc c | cc c | cc c}
	\hline
	\multicolumn{1}{c|}{\multirow{3}{*}{\textbf{Model}}} & \multicolumn{3}{c|}{\textbf{MLQA (7)}} & \multicolumn{3}{c|}{\textbf{XQuAD (12)}} & \multicolumn{3}{c}{\textbf{Belebele (122)}} \\
	\cmidrule(lr){2-3} \cmidrule(lr){4-4} \cmidrule(lr){5-6} \cmidrule(lr){7-7} \cmidrule(lr){8-9} \cmidrule(lr){10-10}
	& \multicolumn{2}{c}{\textbf{nDCG@20}} & \multirow{2}{*}{\textbf{LPR}} & \multicolumn{2}{c}{\textbf{nDCG@20}} & \multirow{2}{*}{\textbf{LPR}} & \multicolumn{2}{c}{\textbf{nDCG@200}} & \multirow{2}{*}{\textbf{LPR}} \\[-0.3em]
	& \small \textbf{Base} & \small \textbf{Lang} & & \small \textbf{Base} & \small \textbf{Lang} & & \small \textbf{Base} & \small \textbf{Lang} & \\
	\hline
	\multicolumn{10}{c}{\textit{\textbf{Dense Retrievers}}} \\
	\hline
	Qwen3-Embedding-8B & \textbf{68.64} & \underline{64.97} & 53.00 & 92.38 & 91.06 & 76.46 & \textbf{74.35} & \textbf{74.41} & 73.22 \\
	\rowcolor{gray!8} Qwen3-Embedding-4B & 66.05 & 62.64 & 52.30 & 91.57 & 90.04 & 73.92 & 62.58 & 62.75 & 65.09 \\
	Qwen3-Embedding-0.6B & 55.16 & 53.79 & 53.65 & 79.44 & 79.55 & 72.78 & 43.09 & 43.61 & 53.82 \\
	\rowcolor{gray!8} llama-embed-nemotron-8b & \underline{68.31} & \textbf{66.82} & 64.55 & 94.37 & \underline{93.54} & 83.91 & \underline{64.68} & \underline{65.12} & 71.10 \\
	llama-nemotron-embed-1b-v2 & 63.61 & 63.23 & 63.99 & \underline{94.56} & \textbf{93.86} & 84.11 & 38.98 & 40.36 & 72.15 \\
	\rowcolor{gray!8} pplx-embed-v1-4b & 67.07 & 62.74 & 46.65 & \textbf{94.94} & 90.24 & 51.51 & 58.12 & 57.05 & 25.59 \\
	pplx-embed-v1-0.6b & 61.54 & 60.11 & 58.06 & 89.57 & 88.17 & 72.11 & 49.31 & 49.24 & 43.09 \\
	\rowcolor{gray!8} jina-embeddings-v5-small & 62.07 & 58.66 & 49.55 & 90.67 & 88.19 & 65.47 & 58.85 & 58.63 & 51.63 \\
	jina-embeddings-v5-nano & 61.25 & 57.79 & 48.42 & 90.13 & 87.90 & 67.18 & 55.81 & 55.73 & 53.26 \\
	\rowcolor{gray!8} jina-embeddings-v3 & 28.72 & 28.24 & 42.00 & 68.96 & 65.69 & 36.65 & 29.20 & 29.04 & 24.98 \\
	gte-Qwen2-7B-instruct & 60.34 & 58.52 & 57.98 & 88.93 & 85.86 & 59.16 & 58.20 & 57.31 & 34.72 \\
	\rowcolor{gray!8} gte-Qwen2-1.5B-instruct & 51.27 & 55.24 & 80.78 & 76.66 & 79.72 & 93.08 & 38.67 & 39.74 & 66.11 \\
	gte-multilingual-base & 54.12 & 52.65 & 51.82 & 83.31 & 80.50 & 53.77 & 62.37 & 60.71 & 18.04 \\
	\rowcolor{gray!8} snowflake-arctic-embed-l-v2.0 & 54.97 & 55.65 & 63.79 & 85.31 & 85.28 & 78.62 & 49.71 & 50.20 & 58.52 \\
	snowflake-arctic-embed-m-v2.0 & 45.14 & 48.87 & 71.93 & 65.56 & 69.93 & 86.41 & 29.48 & 30.96 & 68.23 \\
	\rowcolor{gray!8} multilingual-e5-large & 42.63 & 52.49 & \textbf{96.15} & 69.92 & 75.55 & \textbf{99.92} & 42.40 & 43.55 & 62.76 \\
	multilingual-e5-base & 40.23 & 49.86 & 94.73 & 61.68 & 68.73 & 99.68 & 35.68 & 37.01 & 62.10 \\
	\rowcolor{gray!8} multilingual-e5-small & 35.97 & 46.42 & \underline{95.91} & 51.11 & 60.29 & 99.73 & 23.66 & 24.76 & 44.14 \\
	harrier-oss-v1-0.6b & 43.77 & 53.15 & 95.66 & 62.60 & 69.64 & \underline{99.75} & 33.00 & 35.19 & 81.70 \\
	\rowcolor{gray!8} harrier-oss-v1-270m & 42.48 & 51.08 & 94.17 & 60.61 & 67.79 & 99.56 & 32.55 & 34.71 & 80.99 \\
	voyage-4-nano & 62.54 & 59.66 & 52.86 & 91.89 & 88.46 & 57.84 & 63.43 & 62.59 & 37.41 \\
	\rowcolor{gray!8} embeddinggemma-300m & 61.22 & 60.80 & 67.44 & 86.85 & 87.33 & 86.35 & 49.06 & 49.28 & 53.22 \\
	bge-m3 & 56.65 & 55.43 & 53.79 & 88.06 & 85.94 & 64.42 & 59.41 & 58.81 & 39.25 \\
	\rowcolor{gray!8} nomic-embed-text-v2-moe & 52.89 & 55.03 & 67.59 & 82.36 & 83.37 & 82.01 & 45.12 & 46.08 & 64.86 \\
	granite-embedding-278m & 43.56 & 44.65 & 58.36 & 70.38 & 70.83 & 64.49 & 35.14 & 35.52 & 44.14 \\
	\rowcolor{gray!8} paraphrase-multilingual-MiniLM & 34.92 & 32.65 & 36.88 & 70.55 & 64.98 & 23.47 & 24.13 & 23.63 & 11.63 \\
	LaBSE & 28.41 & 26.85 & 37.53 & 64.70 & 59.99 & 24.03 & 42.18 & 40.88 & 8.96 \\
	\hline
	\multicolumn{10}{c}{\textit{\textbf{Sparse Retrievers}}} \\
	\hline
	opensearch-neural-sparse-v1 & 29.98 & 38.21 & 91.03 & 29.19 & 39.88 & 95.24 & 15.90 & 17.78 & 71.66 \\
	\rowcolor{gray!8} BM25 (XLM-R tokenizer) & 27.92 & 38.23 & 93.68 & 23.31 & 37.77 & 98.75 & 7.27 & 10.27 & \textbf{89.76} \\
    bge-m3-sparse & 33.69 & 42.58 & 87.77 & 33.47 & 45.74 & 96.02 & 19.86 & 22.37 & 74.86 \\
	\hline
	\multicolumn{10}{c}{\textit{\textbf{Late-Interaction Retrievers}}} \\
	\hline
	jina-colbert-v2 & 62.09 & 61.78 & 60.42 & 89.42 & 88.75 & 79.10 & 44.70 & 45.66 & 66.93 \\
	\rowcolor{gray!8} LFM2-ColBERT-350M & 39.48 & 49.00 & 90.73 & 32.74 & 44.02 & 98.41 & 10.06 & 12.44 & \underline{87.75} \\
    bge-m3-colbert & 61.74 & 61.76 & 62.02 & 90.90 & 90.28 & 79.75 & 59.69 & 59.65 & 53.07 \\
	\hline
	\end{tabular}
	}
\end{table*}

\subsection{Main Results}
\label{sec:main-results}

\paragraph{nDCG--LPR Mismatch}
Table~\ref{tab:main-results} reports the main results for all 33 retrievers.
The central observation is that standard language-agnostic retrieval quality does not reliably predict whether a retriever prioritizes query-language passages.
Across the three datasets, the association between standard nDCG and LPR is consistently negative: the Pearson/Spearman correlations are $-0.42$/$-0.35$ on MLQA, $-0.40$/$-0.50$ on XQuAD, and $-0.32$/$-0.33$ on Belebele, with $p<0.05$ for both coefficients on MLQA and XQuAD (Appendix~\ref{app:significance}).
This shows that semantic retrieval quality and query-language preference form distinct behavioral axes.
The Lang-nDCG results further reflect this tension. Models with similar Base nDCG can receive different language-aware scores depending on whether they prioritize query-language passages.
For example, on MLQA, Qwen3-Embedding-8B achieves higher Base nDCG than llama-embed-nemotron-8b, but lower Lang-nDCG, consistent with its much weaker LPR.
Thus, high semantic retrieval quality can coexist with weak query-language preference, where relevant content is retrieved but not necessarily in the query language.

\paragraph{Patterns Across Retrieval Paradigms}
The nDCG--LPR mismatch also appears within each retrieval paradigm; we report these patterns as observations on the evaluated models rather than as general properties of a paradigm.
Dense retrievers show large variation across model families: recent large-scale embedding models such as Qwen3-Embedding and llama-nemotron achieve strong nDCG, whereas the multilingual-e5 family shows much stronger LPR despite lower nDCG.
The evaluated sparse retrievers show a more consistent pattern.
BM25, OpenSearch-NSE, and bge-m3-sparse achieve high LPR, but their standard nDCG is much lower than that of the strongest dense and late-interaction retrievers.
This is consistent with their reliance on lexical overlap, which favors query-language evidence but limits cross-lingual semantic matching.
The late-interaction retrievers span a wider range.
jina-colbert-v2 and bge-m3-colbert achieve competitive semantic retrieval, suggesting that token-level matching can provide fine-grained multilingual relevance signals, whereas LFM2-ColBERT preserves query language more strongly while showing weaker semantic retrieval.

\paragraph{Model Family and Scale Patterns}
LPR also varies with model family and scale, but our observational comparison cannot separate these factors.
Within a family, the effect of scale is dataset-dependent: from Qwen3-Embedding-0.6B to 8B, LPR changes by +3.7 points on XQuAD and $-$0.6 on MLQA, but by +19.4 points on Belebele, and multilingual-e5 shows a similar +18.6-point gain from small to large on Belebele with little change on XQuAD and MLQA.
Cross-family differences are larger, yet families differ simultaneously in architecture, parameter count, tokenizer, pretraining, and fine-tuning data.
One hypothesis consistent with these patterns concerns the language composition of retrieval supervision: Qwen3-Embedding and llama-nemotron construct cross-lingual relevance pairs, where queries and positive passages are written in different languages~\citep{zhang2025qwen3, babakhin2025llamaembednemotron8buniversaltextembedding}, whereas multilingual-e5 constructs query--positive pairs within the same language~\citep{wang2024multilingual}.
A similar contrast holds for late-interaction retrieval: jina-colbert-v2 uses cross-lingual relevance supervision during retrieval fine-tuning~\citep{xiao-etal-2024-jina}, whereas LFM2-ColBERT is initialized from a multilingual model but fine-tuned on English data~\citep{liquidai2025lfm2}.
Establishing a causal role for training data would require a controlled ablation under fixed architecture and scale, which we leave to future work.

\begin{figure*}[t!]
\centering
\includegraphics[width=0.84\textwidth, trim={0 0.2cm 0 0.1cm}, clip]{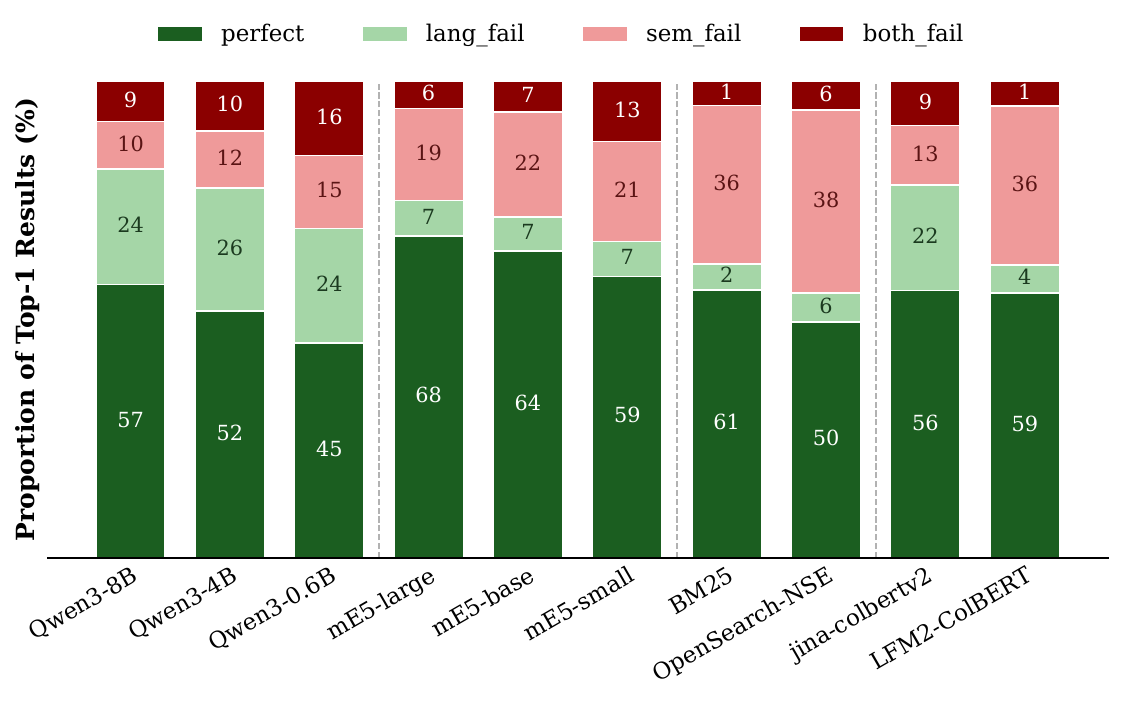}
\caption{
Failure decomposition across representative retrievers.
Each bar partitions the top-ranked result into \texttt{perfect}, \texttt{lang\_fail}, \texttt{sem\_fail}, and \texttt{both\_fail}.
}
\label{fig:scoredist}
\end{figure*}

\subsection{Analysis of Top-1 Ranked Results}
\label{sec:scoredist}

\paragraph{Decomposition Setup}
Figure~\ref{fig:scoredist} decomposes each retriever's top-ranked result into the four outcomes defined in Section~\ref{sec:top1_4way_decomposition}.
The reported proportions are macro-averaged over the three \textsc{MLAIRE} datasets - Belebele, XQuAD, and MLQA.
We include representative models from the main behavioral regimes in Table~\ref{tab:main-results}: Qwen3-Embedding as a semantically strong dense retriever, multilingual-e5 (mE5) as a language-preserving dense retriever, BM25 and OpenSearch-NSE as sparse retrievers, and jina-colbert-v2 and LFM2-ColBERT as late-interaction retrievers.

\paragraph{Semantic Match with Query-Language Mismatch}
Qwen3-Embedding-8B and jina-colbert-v2 frequently retrieve semantically relevant passages at the top rank, as shown by their high combined \texttt{perfect} and \texttt{lang\_fail} rates.
However, a substantial share of these semantic matches are \texttt{lang\_fail}: the model retrieves the correct content, but in a language different from the query.
Standard nDCG counts this outcome as successful because the retrieved content is semantically relevant.
This explains why strong semantic retrieval does not imply strong query-language preference.

\paragraph{Query-Language Retrieval with Semantic Failures}
The opposite pattern appears for mE5 and sparse retrievers.
These models more often preserve the query language, but their non-perfect outcomes are more frequently semantic failures.
In particular, BM25 and OpenSearch-NSE rarely produce \texttt{lang\_fail}, yet they show much larger \texttt{sem\_fail} and \texttt{both\_fail} proportions.
This shows that strong query-language preference does not necessarily imply successful semantic retrieval.
For sparse retrievers, high language preservation can partly reflect lexical anchoring, which favors same-language passages even when they do not contain the correct content.

\paragraph{Grounded versus Spurious Preference}
LPR alone does not reveal whether query-language preference is semantically grounded, so we read it jointly with the decomposition.
Preference is grounded when high LPR co-occurs with a high \texttt{perfect} rate, and spurious when it co-occurs with a high \texttt{sem\_fail} rate, where the top-ranked passage matches the query language but not the content.
On Belebele, for example, BM25 attains the highest LPR (89.76), yet only 49.96\% of its rank-1 passages are \texttt{perfect} and 43.72\% are \texttt{sem\_fail}, whereas harrier-oss-v1-0.6b has a lower LPR (81.70) but a higher \texttt{perfect} rate (60.81\%) and about half the \texttt{sem\_fail} rate (22.36\%).
Table~\ref{tab:lpr-top1} in Appendix~\ref{app:lpr-top1} reports LPR alongside both rates for all 33 retrievers.

% Preamble
% \usepackage{booktabs}
% \usepackage{graphicx}
% \usepackage{multirow}

\begin{table*}[t!]
\centering
\caption{Per-query-language LPR on XQuAD. Resource tiers follow Table~\ref{tab:nllb_resource_classification}}
\label{tab:query_lpr}
\small
\renewcommand{\arraystretch}{1.15}
\setlength{\tabcolsep}{6pt}
\resizebox{\textwidth}{!}{
\begin{tabular}{l ccccccccc ccc | c}
\toprule
\multirow{2}{*}{\textbf{Model}} 
& \multicolumn{9}{c}{\textbf{High-resource}} 
& \multicolumn{3}{c|}{\textbf{Mid-resource}} 
& \multirow{2}{*}{\textbf{AVG}} \\
\cmidrule(lr){2-10}
\cmidrule(lr){11-13}
& \textbf{de} & \textbf{es} & \textbf{en} & \textbf{vi} & \textbf{hi} & \textbf{ar} & \textbf{zh} & \textbf{tr} & \textbf{ru}
& \textbf{ro} & \textbf{th} & \textbf{el}
& \\
\midrule
Qwen3-Embedding-8B   
& 63.61 & 60.67 & 53.28 & 68.24 & 73.11 & 71.01 & 61.93 & 90.67 & 93.70 
& 97.98 & 86.30 & 96.97 & 76.46 \\

Qwen3-Embedding-4B   
& 56.05 & 56.05 & 54.87 & 61.43 & 66.30 & 70.34 & 67.65 & 84.87 & 91.26 
& 87.23 & 94.12 & 97.23 & 73.95 \\

Qwen3-Embedding-0.6B 
& 54.96 & 59.33 & 51.34 & 68.99 & 72.35 & 75.80 & 74.87 & 73.36 & 80.67 
& 81.60 & 93.19 & 86.55 & 72.75 \\

\midrule
multilingual-e5-large
& \textbf{99.92} & \textbf{100.00} & \textbf{99.75} & \textbf{99.92} & \textbf{100.00} & \textbf{99.92} & \textbf{100.00} & \textbf{100.00} & 99.83 
& \textbf{100.00} & 99.92 & \textbf{99.75} & \textbf{99.92} \\

multilingual-e5-base 
& 99.66 & 99.83 & \textbf{99.75} & 99.50 & 99.50 & 99.66 & \textbf{100.00} & 99.66 & 99.75 
& 99.92 & 99.33 & 99.58 & 99.68 \\

multilingual-e5-small
& 99.83 & 99.83 & 99.33 & \textbf{99.92} & 99.83 & 99.75 & \textbf{100.00} & 99.33 & \textbf{99.92} 
& 99.33 & \textbf{100.00} & 99.66 & 99.73 \\

\midrule
BM25                 
& 97.56 & 97.98 & 97.82 & 99.58 & \textbf{100.00} & 99.66 & 99.33 & 97.06 & 98.82 
& 98.32 & 99.58 & 98.66 & 98.70 \\

OpenSearch-NSE
& 89.58 & 92.27 & 91.09 & 97.73 & 99.24 & 98.99 & 98.99 & 95.63 & 98.66 
& 93.03 & 80.67 & 97.73 & 94.47 \\

\midrule
Jina-ColBERT-v2      
& 63.28 & 63.61 & 89.58 & 82.86 & 79.24 & 79.66 & 92.61 & 75.63 & 81.93 
& 66.81 & 88.15 & 75.04 & 78.20 \\

LFM2-ColBERT-350M    
& 96.47 & 97.98 & 91.43 & 99.58 & 99.33 & 99.16 & 99.83 & 99.16 & 99.41 
& 98.82 & 95.97 & 96.72 & 97.82 \\
\bottomrule
\end{tabular}
}
\end{table*}

\subsection{Query-Language Variation in LPR}
\label{sec:per-lang-lpr}

\paragraph{Variation Across Query Languages}
Aggregated LPR can hide substantial variation across query languages, so we report per-query-language LPR on XQuAD in Table~\ref{tab:query_lpr}.
The XQuAD languages in the table include high-resource languages from de to ru and mid-resource languages from ro to el, following the resource tiers in Table~\ref{tab:nllb_resource_classification}.
Among semantically strong retrievers, the Qwen3-Embedding family and jina-colbert-v2 show the clearest language-dependent variation.
For Qwen3-Embedding, LPR is particularly low for several high-resource languages such as German (de), Spanish (es), English (en), Vietnamese (vi), and Chinese (zh), while it is much higher for mid-resource languages.
jina-colbert-v2 also varies substantially across query languages, with low LPR for German (de), Spanish (es), Romanian (ro), and Greek (el), but much higher LPR for English (en), Chinese (zh), and Thai (th).
These results show that query-language preference is not uniform across languages, even within the same model family.

\paragraph{Stable LPR Across Languages}
Other retrievers show more stable LPR patterns.
The multilingual-e5 family, BM25, and LFM2-ColBERT remain close to the LPR ceiling for most XQuAD query languages, while OpenSearch-NSE is generally high but less uniform.
However, high LPR does not by itself indicate strong semantic retrieval, since it can also arise from lexical anchoring or limited cross-lingual matching.
Thus, per-language LPR identifies which query languages are vulnerable, but not where the model retrieves from when query-language preference fails.

\begin{figure}[t]
\centering
\includegraphics[width=\linewidth]{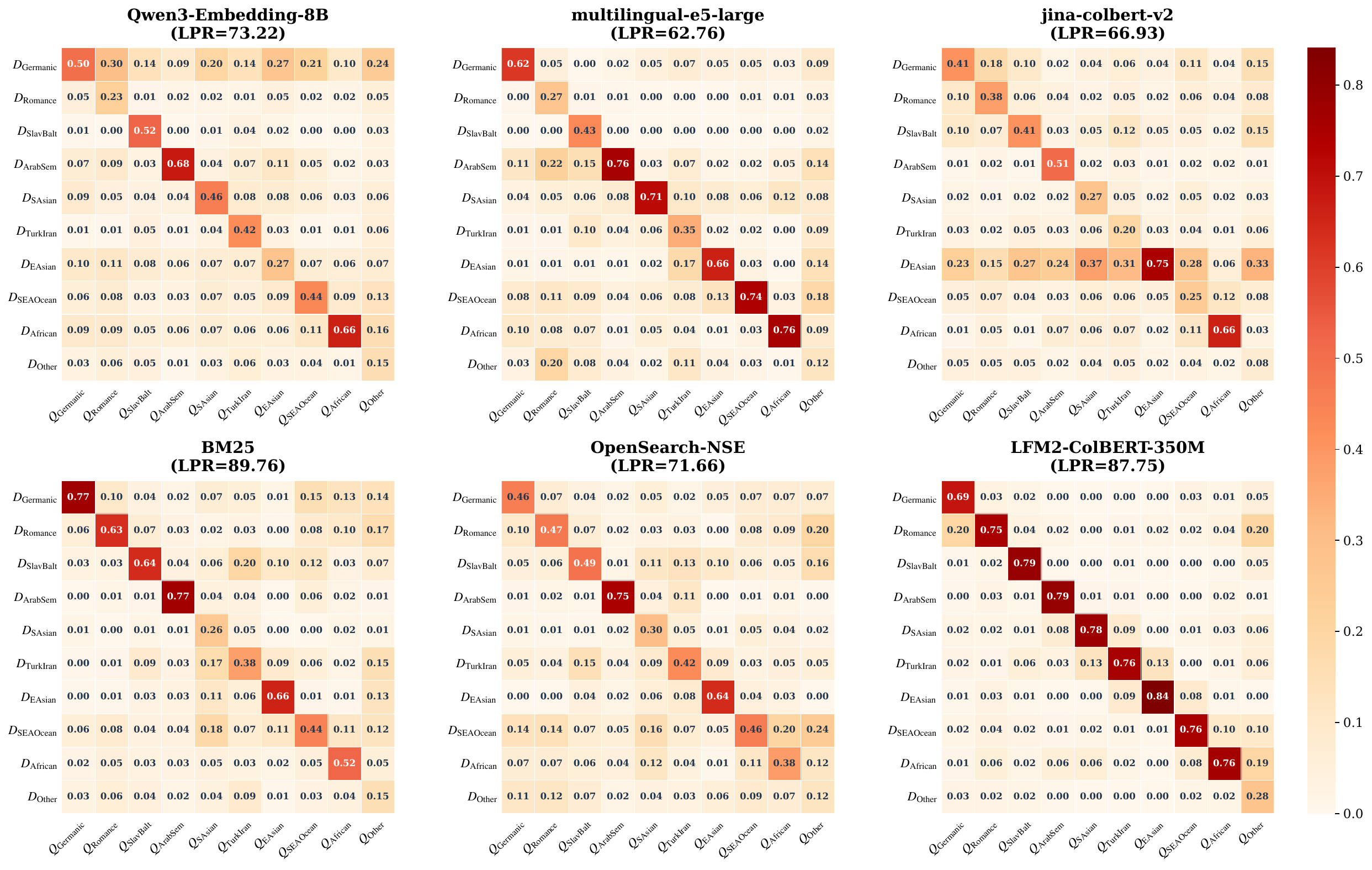}
\caption{
Transition distributions of LPR-failure cases across Belebele macro language groups.
Columns denote the query-language group and rows denote the retrieved document-language group.
Each cell reports the proportion of wrong-language selections assigned to the corresponding document group.
}
\label{fig:resource-gravity}
\end{figure}

\subsection{Directional Query-Language Mismatch by Language Group}
\label{sec:directional-mismatch}

To analyze where query-language preference fails, we group the 122 Belebele languages into ten macro language groups following the classification in Appendix~\ref{app:belebele_macro_groups}.
Figure~\ref{fig:resource-gravity} reports the transition distribution over LPR-failure cases.
Columns denote the query-language group, rows denote the retrieved document-language group, and each cell shows the proportion of non-query-language selections assigned to the corresponding document group.

\paragraph{Diagonal Concentration Across Language Groups}
Non-query-language retrieval is highly structured across macro language groups.
The dominant pattern is diagonal concentration: when retrievers fail to select the query language passage, they often select a semantically equivalent passage from another language in the same macro group.
% This trend is visible across several retrievers, including Qwen3-Embedding-8B, multilingual-e5-large, BM25, and LFM2-ColBERT.
For example, Germanic, Slavic-Baltic, Arabic-Semitic, South Asian, Southeast Asian, and Sub-Saharan African queries frequently retrieve documents from their own macro groups, and this trend is visible across all retrievers.
This indicates that query-language preference failures are not random language confusions, but are structured by language-family, script, or regional affinity.

\paragraph{Model-Specific Directional Patterns}
The strength and direction of this structure differ across retrievers.
Qwen3-Embedding-8B and multilingual-e5-large show clear within-group concentration for many query groups, but also exhibit cross-group attraction, such as Romance queries frequently retrieving Germanic documents.
BM25, OpenSearch-NSE, and LFM2-ColBERT show the strongest diagonal concentration among the displayed models, consistent with their high LPR and stronger tendency to preserve language or script-level similarity.
By contrast, Jina-ColBERT-v2 exhibits a more asymmetric pattern.
Its failures are often attracted toward East Asian documents across many query groups, rather than being explained solely by same-group substitution.
These differences indicate that non-query-language retrieval patterns are model-specific rather than determined by architecture alone.

Overall, the macro-group transition analysis complements per-language LPR.
Per-language LPR identifies which query languages are vulnerable, while Figure~\ref{fig:resource-gravity} shows which document-language groups are selected when query-language preference fails.
The results show that LPR failures are often structured by group-level linguistic affinity and model-specific target-language attraction, rather than by random selection among languages.

\section{Conclusion}
We introduce \textbf{\textsc{mlaire}}, a multilingual language-aware retrieval evaluation protocol that disentangles semantic retrieval quality from query-language preference.
Across 33 retrievers, we show that standard retrieval metrics do not reliably indicate whether a model retrieves evidence in the query language.
Using \textsc{mlaire}, dense, sparse, and late-interaction retrievers exhibit distinct trade-offs: semantically strong models can retrieve relevant evidence in a non-query language, while language-preserving models may lack robust cross-lingual semantic coverage.
Through top-1 failure decomposition, per-language LPR analysis, and macro-group transition analysis, we further show that non-query-language retrieval is often structured.
These findings highlight the need to evaluate multilingual retrievers not only by what they retrieve, but also by whether their retrieval behavior aligns with users' language expectations.

\begin{ack}
This research was supported by Culture, Sports and Tourism R\&D Program, funded by the Ministry of Culture, Sports and Tourism, through the Korea Culture Technology Planning and Evaluation Institute, an affiliated institute of the Korea Creative Content Agency in 2026 (Project Name: Development of an AI Agent Integrating Korean Language Knowledge for Personalized Language Consultation Services, Project Number: RS-2026-25506607, Contribution Rate: 34\%). 
This work was supported by Institute of Information \& communications Technology Planning \& Evaluation (IITP) under the Leading Generative AI Human Resources Development (IITP-2026-RS-2024-00397085) grant funded by the Korea government (MSIT). 
This work was supported by the Institute of Information \& Communications Technology Planning \& Evaluation (IITP)-ICT Creative Consilience Program grant funded by the Korea government (MSIT) (IITP-2026-RS-2020-II201819).
\end{ack}

\bibliographystyle{unsrtnat} % 또는 unsrtnat 등 원하는 스타일 지정
\bibliography{references}

%%%%%%%%%%%%%%%%%%%%%%%%%%%%%%%%%%%%%%%%%%%%%%%%%%%%%%%%%%%%
\clearpage
\appendix
\section*{Appendix}

\appendix

% =============================================================
\section{Limitations}
\label{app:limitations}

\paragraph{Scope of query-language preference}
\textsc{MLAIRE} treats the query-language version of a semantically equivalent passage as the preferred retrieval target when such a passage is available.
This is appropriate for many user-facing search and RAG scenarios, where users expect to read and verify evidence in the language of their query.
However, LPR should not be interpreted as a universal utility measure.
For bilingual users, code-switching contexts, or domains where high-quality information is more abundant in a non-query language, cross-lingual evidence may be equally useful or even preferable.
Therefore, LPR is best read as a utility axis conditional on a need to read or verify evidence in the query language, and LPR and Lang-nDCG should be used as complementary diagnostics rather than replacements for standard semantic retrieval metrics.
We rely on prior studies of multilingual search behavior for this assumption rather than a new user study; validating it with a human preference study across language communities and task types is a direction for future work.

\paragraph{Dataset scope}
Our evaluation is built from Belebele, XQuAD, and MLQA, which provide fully or partially parallel QA-style passages.
This construction creates a controlled setting where semantic relevance can be separated from query-language preference.
The parallel construction is a methodological requirement rather than an incidental choice: without semantically equivalent passages across languages, a low-ranked query-language passage cannot be distinguished from the absence of any query-language version.
Our claims are therefore limited to this controlled QA setting, which may not fully represent open-domain mixed-language corpora, where documents can differ in length, style, topical coverage, translation quality, and native-language authorship.
Extending the protocol to such corpora requires constructing or mining cross-lingual equivalence sets, for example in ad hoc retrieval collections with native relevance judgments across languages and naturally occurring mixed-language document pools, which we leave to future work.

\paragraph{Partial parallelism for MLQA}
Belebele and XQuAD are fully parallel, whereas MLQA is only partially parallel.
As a result, the number of semantically equivalent language alternatives differs across datasets.
This can affect both LPR and Lang-nDCG because the set of competing translations is smaller in MLQA.

\paragraph{Considerations for new metrics}
LPR measures whether the query-language version of the target content receives the highest score among semantically equivalent alternatives.
A model could increase LPR by over-weighting language-identification or surface-form cues after matching the content group, without improving semantic retrieval.
This is why \textsc{MLAIRE} reports LPR together with standard nDCG, Lang-nDCG, and the top-1 4-way decomposition.
Future diagnostic controls could include language-confusable distractors, script-normalized variants, and adversarial same-language passages that share surface cues but not content.

\paragraph{RAG motivation experiment}
The RAG experiment in Section~\ref{sec:background} is intended as a motivating analysis.
It controls semantic content by comparing English and query-language gold passages, but the current presentation does not exhaustively vary retrieval depth, passage chunking, prompt templates, decoding settings, or answer-normalization choices.
These details can affect answer accuracy and language coherence.
Future work should evaluate whether the same pattern holds under retrieved rather than gold passages, multiple generators, and broader decoding configurations.

\paragraph{Fairness and language bias}
Query-language preference overlaps with language fairness in retrieval, but the objectives are not identical.
A model with high LPR may still underserve users if it retrieves same-language but semantically weak evidence.
Conversely, a strong cross-lingual retriever may receive lower LPR even when it benefits bilingual users or users seeking content from other language communities.
Accordingly, \textsc{MLAIRE} should be used to diagnose retrieval behavior, not to enforce monolingual retrieval in all multilingual settings.

% =============================================================
\section{Ethical Considerations}
\label{app:ethical-considerations}

\paragraph{Dataset licenses and reuse}
\textsc{MLAIRE} is constructed from public multilingual datasets.
Belebele is released under CC-BY-SA 4.0, XQuAD is distributed with a CC-BY-SA 4.0 license file, and MLQA states that its dataset, derived from Wikipedia paragraphs, is licensed under CC-BY-SA 3.0.
The \textsc{MLAIRE} versions of these datasets are derivatives and are redistributed under the same licenses as their sources.
% Table~\ref{tab:dataset-licenses} summarizes the licenses and reuse considerations.
Users of \textsc{MLAIRE} should follow the attribution and share-alike requirements of the underlying datasets.

% \begin{table}[h]
% \centering
% \caption{Licenses and reuse terms for the datasets used in \textsc{mlaire}.}
% \label{tab:dataset-licenses}
% \small
% \begin{tabular}{p{0.18\linewidth}p{0.22\linewidth}p{0.50\linewidth}}
% \toprule
% \textbf{Dataset} & \textbf{License} & \textbf{Reuse considerations} \\
% \midrule
% Belebele 
% & CC-BY-SA 4.0 
% & Requires attribution and share-alike compliance for redistributed derivatives. \\
% XQuAD 
% & CC-BY-SA 4.0 
% & Requires attribution and share-alike compliance; built from SQuAD v1.1 and professional translations. \\
% MLQA 
% & CC-BY-SA 3.0 
% & Derived from Wikipedia paragraphs; requires attribution and share-alike compliance. \\
% \bottomrule
% \end{tabular}
% \end{table}

\paragraph{Representativeness}
Although the benchmark covers many languages, coverage is still uneven across language families, scripts, and resource levels.
High-resource languages and translated benchmark content may be overrepresented relative to naturally occurring multilingual search scenarios.
Results should therefore be interpreted as controlled measurements of retrieval behavior, not as comprehensive evidence of equitable performance across all linguistic communities.

\paragraph{Responsible use of LPR}
LPR should not be used as a standalone deployment objective.
It is most appropriate when the application expects users to read or verify evidence in the same language as their query.
It may be less appropriate for bilingual users, code-switching users, cross-lingual research workflows, or settings where the best available evidence is naturally written in another language.
Operational use should combine LPR with semantic retrieval metrics and application-specific user needs.

\paragraph{Fairness trade-offs}
Optimizing for query-language preference can improve accessibility for users who need same-language evidence, but it can also penalize systems that intentionally retrieve cross-lingual evidence for broader coverage.
Conversely, high LPR with low semantic retrieval quality can still harm users by returning readable but incorrect or irrelevant evidence.
The intended use of \textsc{MLAIRE} is diagnostic: it helps separate semantic retrieval quality from language preference so that developers can understand trade-offs rather than optimize a single metric blindly.

\paragraph{Positive impact} 
\textsc{MLAIRE} can help developers diagnose whether multilingual retrievers provide evidence in a language users can read and verify, improving accessibility and transparency in multilingual search and RAG systems.

\paragraph{Potential negative impact} 
If LPR is optimized in isolation, systems may over-prioritize same-language passages even when cross-lingual evidence is more complete or reliable. We therefore recommend using LPR only together with semantic retrieval metrics and application-specific user needs.

% =============================================================
\section{Instruction Templates}
\label{app:instruction-templates}

Table~\ref{tab:instruction-templates} lists the query and passage prefixes used for each dense retriever, grouped by model family.
When a model exposes MTEB-style task prompts, we use those prompts. Otherwise we fall back to the prefixes below, which match each model's original release recipe.
For sparse and multi-vector retrievers, no instruction template is applied.

% keep your existing Table~\ref{tab:instruction-templates} here
\begin{table}[ht]
\centering
\caption{Query and passage prefixes used for dense retrievers.}
\label{tab:instruction-templates}
\small
\renewcommand{\arraystretch}{1.18}
\setlength{\tabcolsep}{5pt}
\begin{tabularx}{\linewidth}{
>{\raggedright\arraybackslash}p{0.30\linewidth}
>{\raggedright\arraybackslash}X
>{\raggedright\arraybackslash}p{0.25\linewidth}
}
\toprule
\textbf{Model family} & \textbf{Query prefix} & \textbf{Passage prefix} \\
\midrule

\rowcolor{gray!8}
multilingual-e5
& query: 
& passage:  \\

llama-nemotron-embed-1b-v2
& query: 
& passage:  \\

\rowcolor{gray!8}
snowflake-arctic-embed-l-v2.0
& query: 
& -- \\

nomic-embed-text-v2-moe
& search\_query: 
& search\_document:  \\

\rowcolor{gray!8}
jina-embeddings-v3
& Represent the query for retrieving evidence documents:
& Represent the document for retrieval:  \\

voyage-4-nano
& Represent the query for retrieving supporting documents: 
& Represent the document for retrieval:  \\

\rowcolor{gray!8}
jina-embeddings-v5
& \texttt{Query: }
& \texttt{Document: } \\

embeddinggemma-300m
& task: search result | query: 
& title: none | text:  \\

\rowcolor{gray!8}
harrier-oss-v1 
& Instruct: Given a question, retrieve passages that answer the question\textbackslash n Query: 
& -- \\

Qwen3-Embedding 
& Instruct: Given a question, retrieve passages that answer the question\textbackslash n Query: 
& -- \\

\rowcolor{gray!8}
llama-embed-nemotron-8b
& Instruct: Given a question, retrieve passages that answer the question\textbackslash n Query: 
& -- \\

\rowcolor{gray!8}
gte-Qwen2-instruct
& Instruct: Given a web search query, retrieve relevant passages that answer the query \textbackslash n Query: 
& -- \\

bge-m3
& --
& -- \\

\rowcolor{gray!8}
gte-multilingual-base
& --
& -- \\

pplx-embed-v1
& --
& -- \\

\rowcolor{gray!8}
granite-embedding-278m
& --
& -- \\

LaBSE
& --
& -- \\

\rowcolor{gray!8}
paraphrase-multilingual-MiniLM
& --
& -- \\

\bottomrule
\end{tabularx}
\end{table}

% =============================================================
\section{Infrastructure and Reproducibility}
\label{app:infrastructure}

\paragraph{Hardware}
All experiments were run on a single workstation with four NVIDIA RTX A6000 (48\,GB) GPUs and 1\,TB of system RAM.

\paragraph{Software environments}
We use Python 3.10 and PyTorch 2.8 with CUDA 12.1.
Sparse and multi-vector retrievers depend on distinct indexing toolchains (\texttt{bm25s}, \texttt{pylate}, \texttt{FlagEmbedding}).
To avoid dependency collisions while keeping the evaluator itself identical, we partition the retriever pool across three isolated virtual environments with pinned requirements; all three import the same evaluation core.

% \paragraph{Determinism}
% All model weights are fetched at fixed HuggingFace revisions.
% Encoding is performed with deterministic CUDA kernels (\texttt{torch.backends.cudnn.deterministic=True}, \texttt{torch.use\_deterministic\_algorithms(True)}), and random seeds are fixed at 42 for all stochastic components of the pipeline.
% Given these controls, re-running the evaluation produces bitwise-identical retrieval scores.

\paragraph{Release artifacts}
We release the evaluation code at \url{https://github.com/yjoonjang/MLAIRE} under the MIT License, including the benchmark construction scripts, retriever adapters, the evaluation core, the LPR pipeline, and the setup scripts for the three virtual environments.
The three datasets are released on Hugging Face at \url{https://huggingface.co/collections/yjoonjang/mlaire} under the licenses of their sources, together with dataset cards and Croissant metadata with Responsible AI fields.
New retrievers can be evaluated by supplying a retrieval function to the shared evaluator, without modifying the benchmark itself.
\paragraph{Runtime} End-to-end evaluation time varies by retriever architecture and dataset size. Dense retrievers require corpus encoding followed by top-$k$ retrieval, sparse retrievers require indexing with their respective sparse representations, and late-interaction models require PLAID indexing. In total, the reported experiments required approximately 50 GPU-hours and under 1 CPU-hour (BM25 baseline). The largest cost comes from encoding/indexing the Belebele 122-language corpus across all retrievers.

% =============================================================
\section{Language Classification}
\label{app:language-classification}

\paragraph{Resource-level classification}
For analyses that require language resource levels, we primarily follow the resource classes provided with SIB-200~\cite{adelani2024sib}, which is based on FLORES-200~\cite{costa2022no} and includes Joshi-style resource categories.
We map classes 0--2 to low-resource, class 3 to mid-resource, and classes 4--5 to high-resource languages.
For languages or script variants without a Joshi-style class, we use the NLLB-200~\cite{costa2022no} high/low-resource distinction as a fallback, where low-resource languages are defined by having fewer than 1M publicly available deduplicated bitext sentence pairs.
The resulting resource-level mapping is reported in Table~\ref{tab:nllb_resource_classification}.

\begin{table*}[ht]
\footnotesize
\centering
\caption{Language classification based on resource levels}
\label{tab:nllb_resource_classification}
\renewcommand{\arraystretch}{1.5} % slightly wider row spacing for readability
\setlength{\aboverulesep}{0pt} % remove booktabs padding so the vertical rule is continuous
\setlength{\belowrulesep}{0pt}
\begin{tabular}{l|p{0.8\textwidth}}
\toprule
\textbf{Resource Level} & \textbf{FLORES-200 language codes} \\
\midrule

\textbf{High} & 
arb\_Arab, cat\_Latn, ces\_Latn, deu\_Latn, eng\_Latn, eus\_Latn, fin\_Latn, fra\_Latn, hin\_Deva, hin\_Latn, hrv\_Latn, hun\_Latn, ita\_Latn, jpn\_Jpan, kor\_Hang, nld\_Latn, pes\_Arab, pol\_Latn, por\_Latn, rus\_Cyrl, spa\_Latn, srp\_Cyrl, swe\_Latn, tur\_Latn, vie\_Latn, zho\_Hans \\

\midrule

\textbf{Mid} & 
afr\_Latn, arz\_Arab, ben\_Beng, ben\_Latn, bul\_Cyrl, ceb\_Latn, dan\_Latn, ell\_Grek, est\_Latn, heb\_Hebr, ind\_Latn, kat\_Geor, kaz\_Cyrl, lit\_Latn, lvs\_Latn, ron\_Latn, slk\_Latn, slv\_Latn, tam\_Taml, tgl\_Latn, tha\_Thai, ukr\_Cyrl, urd\_Arab, urd\_Latn, uzn\_Latn, zsm\_Latn \\

\midrule

\textbf{Low} & 
acm\_Arab, als\_Latn, amh\_Ethi, apc\_Arab, arb\_Latn, ars\_Arab, ary\_Arab, asm\_Beng, azj\_Latn, bam\_Latn, bod\_Tibt, ckb\_Arab, fuv\_Latn, gaz\_Latn, grn\_Latn, guj\_Gujr, hat\_Latn, hau\_Latn, hye\_Armn, ibo\_Latn, ilo\_Latn, isl\_Latn, jav\_Latn, kac\_Latn, kan\_Knda, kea\_Latn, khk\_Cyrl, khm\_Khmr, kin\_Latn, kir\_Cyrl, lao\_Laoo, lin\_Latn, lug\_Latn, luo\_Latn, mal\_Mlym, mar\_Deva, mkd\_Cyrl, mlt\_Latn, mri\_Latn, mya\_Mymr, nob\_Latn, npi\_Deva, npi\_Latn, nso\_Latn, nya\_Latn, ory\_Orya, pan\_Guru, pbt\_Arab, plt\_Latn, shn\_Mymr, sin\_Latn, sin\_Sinh, sna\_Latn, snd\_Arab, som\_Latn, sot\_Latn, ssw\_Latn, sun\_Latn, swh\_Latn, tel\_Telu, tgk\_Cyrl, tir\_Ethi, tsn\_Latn, tso\_Latn, war\_Latn, wol\_Latn, xho\_Latn, yor\_Latn, zho\_Hant, zul\_Latn \\

\bottomrule
\end{tabular}
\end{table*}

\begin{table*}[ht]
\centering
\caption{Macro language groups used for the Belebele directional analysis. The short labels correspond to the group labels shown in Figure~\ref{fig:resource-gravity}.}
\label{tab:belebele_macro_groups}
\footnotesize
\renewcommand{\arraystretch}{1.25}
\setlength{\tabcolsep}{5pt}
\begin{tabularx}{\textwidth}{l|>{\raggedright\arraybackslash}p{0.25\textwidth}|>{\raggedright\arraybackslash}X}
\noalign{\hrule height 0.8pt} % thick top rule that keeps vertical rules connected
\textbf{Short label} & \textbf{Macro group} & \textbf{FLORES-200 language codes} \\
\hline
\texttt{Germ}
& Germanic / Anglophone-North European
& afr\_Latn, dan\_Latn, deu\_Latn, eng\_Latn, isl\_Latn, nld\_Latn, nob\_Latn, swe\_Latn \\
\hline
\texttt{Rom}
& Romance / Latin-European + Creole
& cat\_Latn, fra\_Latn, ita\_Latn, por\_Latn, ron\_Latn, spa\_Latn, hat\_Latn, kea\_Latn \\
\hline
\texttt{Slav}
& Slavic-Baltic / Eastern European
& bul\_Cyrl, ces\_Latn, hrv\_Latn, lit\_Latn, lvs\_Latn, mkd\_Cyrl, pol\_Latn, rus\_Cyrl, slk\_Latn, slv\_Latn, srp\_Cyrl, ukr\_Cyrl \\
\hline
\texttt{Arab}
& Arabic-Semitic / MENA
& acm\_Arab, apc\_Arab, arb\_Arab, arb\_Latn, ars\_Arab, ary\_Arab, arz\_Arab, heb\_Hebr, mlt\_Latn \\
\hline
\texttt{SAsia}
& South Asian / Indo-Aryan-Dravidian
& asm\_Beng, ben\_Beng, ben\_Latn, guj\_Gujr, hin\_Deva, hin\_Latn, kan\_Knda, mal\_Mlym, mar\_Deva, npi\_Deva, npi\_Latn, ory\_Orya, pan\_Guru, sin\_Latn, sin\_Sinh, snd\_Arab, tam\_Taml, tel\_Telu, urd\_Arab, urd\_Latn \\
\hline
\texttt{TrIr}
& Iranian-Turkic / Central-West Asian
& azj\_Latn, ckb\_Arab, kaz\_Cyrl, kir\_Cyrl, pbt\_Arab, pes\_Arab, tgk\_Cyrl, tur\_Latn, uzn\_Latn \\
\hline
\texttt{EAsia}
& East Asian / CJK-Mongolic-Tibetic
& bod\_Tibt, zho\_Hans, zho\_Hant, jpn\_Jpan, kor\_Hang, khk\_Cyrl \\
\hline
\texttt{SEA}
& Southeast Asian / Austronesian-Oceanic
& ceb\_Latn, ilo\_Latn, ind\_Latn, jav\_Latn, kac\_Latn, khm\_Khmr, lao\_Laoo, mri\_Latn, mya\_Mymr, plt\_Latn, shn\_Mymr, sun\_Latn, tgl\_Latn, tha\_Thai, vie\_Latn, war\_Latn, zsm\_Latn \\
\hline
\texttt{Afr}
& Sub-Saharan African
& amh\_Ethi, bam\_Latn, fuv\_Latn, gaz\_Latn, hau\_Latn, ibo\_Latn, kin\_Latn, lin\_Latn, lug\_Latn, luo\_Latn, nso\_Latn, nya\_Latn, sna\_Latn, som\_Latn, sot\_Latn, ssw\_Latn, swh\_Latn, tir\_Ethi, tsn\_Latn, tso\_Latn, wol\_Latn, xho\_Latn, yor\_Latn, zul\_Latn \\
\hline
\texttt{Oth}
& Other Europe / Caucasus / Americas
& als\_Latn, ell\_Grek, est\_Latn, eus\_Latn, fin\_Latn, grn\_Latn, hun\_Latn, hye\_Armn, kat\_Geor \\
\noalign{\hrule height 0.8pt} % thick bottom rule
\end{tabularx}
\end{table*}

\paragraph{Macro language groups}
\label{app:belebele_macro_groups}
For the directional analysis in Section~\ref{sec:directional-mismatch}, we group the 122 Belebele language variants into ten macro language groups.
This grouping is not intended as a strict linguistic taxonomy.
Rather, it is an analysis convenience that aggregates languages by broad family, script, and regional affinities so that group-level non-query-language retrieval patterns can be visualized.
Table~\ref{tab:belebele_macro_groups} lists the full mapping and the short labels used in Figure~\ref{fig:resource-gravity}.

\section{Recall and Lang-Recall}
\label{app:recall-comparison}

\begin{table*}[t]
	\centering
	\caption{Recall, Lang-Recall, and LPR across the three datasets for all 33 retrievers. MLQA (7 languages) and XQuAD (12 languages) are evaluated at $k{=}20$; Belebele (122 languages) is evaluated at $k{=}200$. All values are percentages (raw scores $\times 100$). \textbf{Recall} is standard Recall@$k$ over the full multilingual relevant group. \textbf{Lang-Recall} restricts the relevant set to the query-language version of the target content and measures whether that single passage appears within the top-$k$. Per column, the best value is \textbf{bold} and the second best is \underline{underlined}.}
	\label{tab:recall-comparison}
	\renewcommand{\arraystretch}{1.3}
	\setlength{\tabcolsep}{8pt}
	\resizebox{\textwidth}{!}{
	\begin{tabular}{l | cc c | cc c | cc c}
	\hline
	\multicolumn{1}{c|}{\multirow{3}{*}{\textbf{Model}}} & \multicolumn{3}{c|}{\textbf{MLQA (7)}} & \multicolumn{3}{c|}{\textbf{XQuAD (12)}} & \multicolumn{3}{c}{\textbf{Belebele (122)}} \\
	\cmidrule(lr){2-3} \cmidrule(lr){4-4} \cmidrule(lr){5-6} \cmidrule(lr){7-7} \cmidrule(lr){8-9} \cmidrule(lr){10-10}
	& \multicolumn{2}{c}{\textbf{Recall@20}} & \multirow{2}{*}{\textbf{LPR}} & \multicolumn{2}{c}{\textbf{Recall@20}} & \multirow{2}{*}{\textbf{LPR}} & \multicolumn{2}{c}{\textbf{Recall@200}} & \multirow{2}{*}{\textbf{LPR}} \\[-0.3em]
	& \small \textbf{Base} & \small \textbf{Lang} & & \small \textbf{Base} & \small \textbf{Lang} & & \small \textbf{Base} & \small \textbf{Lang} & \\
	\hline
	\multicolumn{10}{c}{\textit{\textbf{Dense Retrievers}}} \\
	\hline
	Qwen3-Embedding-8B & \textbf{76.33} & 80.87 & 53.00 & 93.84 & 99.20 & 76.46 & \textbf{73.76} & 96.22 & 73.22 \\
	\rowcolor{gray!8} Qwen3-Embedding-4B & 73.76 & 78.48 & 52.30 & 92.99 & 98.78 & 73.92 & 61.60 & 92.81 & 65.09 \\
	Qwen3-Embedding-0.6B & 62.28 & 70.57 & 53.65 & 80.46 & 96.33 & 72.78 & 41.02 & 85.38 & 53.82 \\
	\rowcolor{gray!8} llama-embed-nemotron-8b & \underline{75.71} & 83.34 & 64.55 & \underline{95.62} & 99.66 & 83.91 & \underline{63.34} & \textbf{96.54} & 71.10 \\
	llama-nemotron-embed-1b-v2 & 70.67 & 80.84 & 63.99 & 95.53 & 99.75 & 84.11 & 35.75 & 93.55 & 72.15 \\
	\rowcolor{gray!8} pplx-embed-v1-4b & 74.43 & 78.24 & 46.65 & \textbf{95.75} & 98.88 & 51.51 & 56.67 & 79.52 & 25.59 \\
	pplx-embed-v1-0.6b & 68.57 & 76.76 & 58.06 & 90.54 & 98.47 & 72.11 & 47.21 & 82.27 & 43.09 \\
	\rowcolor{gray!8} jina-embeddings-v5-small & 69.96 & 74.67 & 49.55 & 92.06 & 97.73 & 65.47 & 57.77 & 89.40 & 51.63 \\
	jina-embeddings-v5-nano & 69.14 & 73.59 & 48.42 & 91.53 & 97.69 & 67.18 & 54.69 & 88.00 & 53.26 \\
	\rowcolor{gray!8} jina-embeddings-v3 & 33.96 & 42.56 & 42.00 & 69.82 & 84.86 & 36.65 & 26.33 & 59.02 & 24.98 \\
	gte-Qwen2-7B-instruct & 68.04 & 75.43 & 57.98 & 90.14 & 96.55 & 59.16 & 56.98 & 82.72 & 34.72 \\
	\rowcolor{gray!8} gte-Qwen2-1.5B-instruct & 56.81 & 79.26 & 80.78 & 78.12 & 98.82 & 93.08 & 36.21 & 87.67 & 66.11 \\
	gte-multilingual-base & 61.63 & 70.02 & 51.82 & 84.38 & 94.79 & 53.77 & 62.24 & 77.79 & 18.04 \\
	\rowcolor{gray!8} snowflake-arctic-embed-l-v2.0 & 60.97 & 74.55 & 63.79 & 86.63 & 97.76 & 78.62 & 47.14 & 90.60 & 58.52 \\
	snowflake-arctic-embed-m-v2.0 & 48.60 & 72.63 & 71.93 & 64.73 & 97.21 & 86.41 & 26.51 & 88.63 & 68.23 \\
	\rowcolor{gray!8} multilingual-e5-large & 43.97 & \underline{85.17} & \textbf{96.15} & 72.59 & \textbf{99.87} & \textbf{99.92} & 40.21 & 88.56 & 62.76 \\
	multilingual-e5-base & 40.96 & 82.72 & 94.73 & 62.05 & 99.78 & 99.68 & 33.44 & 86.43 & 62.10 \\
	\rowcolor{gray!8} multilingual-e5-small & 35.38 & 80.95 & \underline{95.91} & 49.18 & 99.61 & 99.73 & 21.43 & 67.73 & 44.14 \\
	harrier-oss-v1-0.6b & 45.67 & \textbf{85.24} & 95.66 & 63.76 & \underline{99.82} & \underline{99.75} & 30.73 & \underline{96.24} & 81.70 \\
	\rowcolor{gray!8} harrier-oss-v1-270m & 44.99 & 82.70 & 94.17 & 61.10 & 99.67 & 99.56 & 30.29 & 96.23 & 80.99 \\
	voyage-4-nano & 70.57 & 76.59 & 52.86 & 93.05 & 98.16 & 57.84 & 62.20 & 90.91 & 37.41 \\
	\rowcolor{gray!8} embeddinggemma-300m & 68.31 & 78.53 & 67.44 & 88.46 & 98.63 & 86.35 & 47.21 & 86.25 & 53.22 \\
	bge-m3 & 63.29 & 72.50 & 53.79 & 89.18 & 97.32 & 64.42 & 57.84 & 88.21 & 39.25 \\
	\rowcolor{gray!8} nomic-embed-text-v2-moe & 58.40 & 76.15 & 67.59 & 83.26 & 98.07 & 82.01 & 42.27 & 92.11 & 64.86 \\
	granite-embedding-278m & 49.05 & 64.35 & 58.36 & 71.24 & 92.88 & 64.49 & 32.96 & 79.16 & 44.14 \\
	\rowcolor{gray!8} paraphrase-multilingual-MiniLM & 41.60 & 46.72 & 36.88 & 71.94 & 80.71 & 23.47 & 21.86 & 47.91 & 11.63 \\
	LaBSE & 33.38 & 39.99 & 37.53 & 64.17 & 77.79 & 24.03 & 40.86 & 60.89 & 8.96 \\
	\hline
	\multicolumn{10}{c}{\textit{\textbf{Sparse Retrievers}}} \\
	\hline
	opensearch-neural-sparse-v1 & 29.48 & 69.51 & 91.03 & 22.47 & 90.65 & 95.24 & 13.71 & 83.98 & 71.66 \\
	\rowcolor{gray!8} BM25 (XLM-R tokenizer) & 25.73 & 72.98 & 93.68 & 13.94 & 98.56 & 98.75 & 5.22 & 92.82 & \textbf{89.76} \\
	bge-m3-sparse & 32.19 & 75.36 & 87.77 & 25.11 & 98.56 & 96.02 & 15.81 & 91.54 & 74.86 \\
	\hline
	\multicolumn{10}{c}{\textit{\textbf{Late-Interaction Retrievers}}} \\
	\hline
	jina-colbert-v2 & 66.04 & 77.31 & 60.42 & 89.86 & 97.42 & 79.10 & 41.80 & 91.93 & 66.93 \\
	\rowcolor{gray!8} LFM2-ColBERT-350M & 37.31 & 80.29 & 90.73 & 25.49 & 93.89 & 98.41 & 7.88 & 85.10 & \underline{87.75} \\
	bge-m3-colbert & 67.78 & 80.13 & 62.02 & 91.73 & 99.17 & 79.75 & 57.54 & 92.15 & 53.07 \\
	\hline
	\end{tabular}
	}
\end{table*}

Table~\ref{tab:recall-comparison} reports standard Recall and Lang-Recall at the same evaluation depths used in Table~\ref{tab:main-results} 
(\(k{=}20\) for MLQA and XQuAD, and \(k{=}200\) for Belebele).
Recall@\(k\) measures how many semantically relevant passages are retrieved within the top \(k\), regardless of language.
Lang-Recall@\(k\) restricts the relevant set to the query-language version of the target content and measures whether that version appears within the top \(k\).
Because the two metrics use different relevant sets, they capture different aspects of retrieval behavior.
A retriever can achieve high standard Recall by retrieving many cross-lingual relevant passages while missing the query-language version, or achieve high Lang-Recall while retrieving only a small portion of the full multilingual relevant set.

\paragraph{High-LPR retrievers show higher Lang-Recall}
Retrievers with high LPR, including the multilingual-e5 and harrier-oss families, the sparse retrievers, and LFM2-ColBERT-350M, generally show higher Lang-Recall than standard Recall.
For example, multilingual-e5-large obtains Lang-Recall@20 of 85.17 on MLQA and 99.87 on XQuAD, compared with standard Recall@20 of 43.97 and 72.59.
BM25 shows an even larger gap on XQuAD, with Lang-Recall@20 of 98.56 compared with standard Recall@20 of 13.94.
These results indicate that these retrievers often recover the query-language version of the target content, while retrieving fewer of the cross-lingual alternatives.
This pattern is consistent with their high LPR, but reflects a different top-\(k\) perspective: Lang-Recall measures whether the query-language version is recovered within the retrieved set, whereas LPR measures which language version is preferred within the target content group.

\paragraph{Lower-LPR retrievers show smaller gaps}
Retrievers with lower or more moderate LPR, such as pplx-embed-v1-4b, gte-multilingual-base, jina-embeddings-v3, paraphrase-multilingual-MiniLM, and LaBSE, show smaller gaps between Recall and Lang-Recall.
For instance, on XQuAD, pplx-embed-v1-4b obtains Recall@20 of 95.75 and Lang-Recall@20 of 98.88.
On Belebele, the same model obtains Recall@200 of 56.67 and Lang-Recall@200 of 79.52.
These results suggest that such retrievers distribute retrieval capacity more broadly across semantically equivalent passages in multiple languages, rather than concentrating as strongly on the query-language version.
Accordingly, their standard Recall and Lang-Recall tend to move more closely together.

\paragraph{Largest gap in Belebele}
The difference between Recall and Lang-Recall is most pronounced on Belebele.
Because each Belebele query can have up to 122 semantically relevant language variants, standard Recall has a much larger relevant set than in MLQA or XQuAD.
A retriever that concentrates its top-\(k\) results on a subset of languages can therefore achieve high Lang-Recall while still obtaining relatively low standard Recall.
For example, harrier-oss-v1-0.6b reaches Lang-Recall@200 of 96.24 on Belebele, while its standard Recall@200 is 30.73.
This makes Belebele especially useful for separating query-language recovery from broader cross-lingual coverage.

% =============================================================
\section{Statistical Significance}
\label{app:significance}

\begin{wraptable}{r}{0.48\textwidth}
\vspace{-1em} % pull the table up to align with the first line of text
\centering
\caption{Correlation between nDCG and LPR across the 33 retrievers ($r$: Pearson, $\rho$: Spearman), with Fisher-$z$ 95\% confidence intervals.}
\label{tab:corr-significance}
\footnotesize
\setlength{\tabcolsep}{4pt}
\begin{tabular}{l c r c c}
\toprule
Dataset & Coef. & Value & 95\% CI & $p$ \\
\midrule
\multirow{2}{*}{MLQA}     & $r$    & -0.42 & [-0.67, -0.09]           & 0.015 \\
                          & $\rho$ & -0.35 & [-0.62, -0.01]           & 0.044 \\
\midrule
\multirow{2}{*}{XQuAD}    & $r$    & -0.40 & [-0.65, -0.07]           & 0.021 \\
                          & $\rho$ & -0.50 & [-0.72, -0.18]           & 0.003 \\
\midrule
\multirow{2}{*}{Belebele} & $r$    & -0.32 & [-0.59, \phantom{-}0.03] & 0.074 \\
                          & $\rho$ & -0.33 & [-0.61, \phantom{-}0.01] & 0.059 \\
\bottomrule
\end{tabular}
\vspace{-1em} % tighten the gap between the table and the wrapped text below
\end{wraptable}

\paragraph{Correlation between nDCG and LPR}
Table~\ref{tab:corr-significance} tests the association between standard nDCG and LPR across the 33 retrievers, using the Base nDCG and LPR values in Table~\ref{tab:main-results}.
We compute exact $p$-values with \texttt{scipy.stats.pearsonr} and \texttt{spearmanr}, and 95\% confidence intervals with the Fisher $z$-transformation.
All six coefficients are negative.
Both coefficients are significant at $p<0.05$ on MLQA and XQuAD, and both are negative but not significant on Belebele.

\paragraph{Uncertainty of per-model LPR}
LPR is a proportion over queries, so we quantify its uncertainty with Wilson 95\% intervals.
Treating queries as independent, the largest half-width over all 33 retrievers is 0.48, 0.82, and 0.30 points on MLQA, XQuAD, and Belebele, respectively ($n$ = 42{,}245, 14{,}280, and 109{,}800 queries).
Queries that share a content group are not independent, since they are translations of one another or questions about the same passage.
We therefore also report a conservative bound for the extreme case in which all queries in a content group share the same outcome, using the corresponding effective sample size $(\sum_g m_g)^2 / \sum_g m_g^2$, where $m_g$ is the number of queries in group $g$.
Under this bound, the largest half-width becomes 1.12, 6.81, and 4.50 points.
These intervals remain far narrower than the cross-model LPR ranges (36.88--96.15 on MLQA, 23.47--99.92 on XQuAD, and 8.96--89.76 on Belebele), so the large LPR differences discussed in Section~\ref{sec:results} are not attributable to sampling variation, although retrievers whose LPR differs by only a few points on XQuAD or Belebele should not be strictly ranked.
Per-query-language LPR on XQuAD (Table~\ref{tab:query_lpr}) is computed over 1{,}190 queries per language, for which the query-level Wilson half-width is at most 2.84 points.

% =============================================================
\section{LPR and Top-1 Decomposition}
\label{app:lpr-top1}

Table~\ref{tab:lpr-top1} reports LPR together with the \texttt{perfect} and \texttt{sem\_fail} rates of the top-1 decomposition for all 33 retrievers, following the reading in Section~\ref{sec:scoredist}.
High LPR accompanied by a high \texttt{perfect} rate indicates semantically grounded query-language preference, whereas high LPR accompanied by a high \texttt{sem\_fail} rate indicates that the query language is often preserved by a same-language passage with the wrong content.

\begin{table*}[t]
	\centering
	\caption{LPR read jointly with the top-1 decomposition for all 33 retrievers. For each dataset we report LPR, the share of queries whose rank-1 passage is relevant and in the query language (\texttt{perfect}), and the share whose rank-1 passage is in the query language but irrelevant (\texttt{sem\_fail}). MLQA and XQuAD use all of their languages; Belebele uses all 122 languages. All values are percentages.}
	\label{tab:lpr-top1}
	\renewcommand{\arraystretch}{1.3}
	\setlength{\tabcolsep}{8pt}
	\resizebox{\textwidth}{!}{
	\begin{tabular}{l | ccc | ccc | ccc}
	\hline
	\multicolumn{1}{c|}{\multirow{2}{*}{\textbf{Model}}} & \multicolumn{3}{c|}{\textbf{MLQA (7)}} & \multicolumn{3}{c|}{\textbf{XQuAD (12)}} & \multicolumn{3}{c}{\textbf{Belebele (122)}} \\
	& \small \textbf{LPR} & \small \texttt{perfect} & \small \texttt{sem\_fail} & \small \textbf{LPR} & \small \texttt{perfect} & \small \texttt{sem\_fail} & \small \textbf{LPR} & \small \texttt{perfect} & \small \texttt{sem\_fail} \\
	\hline
	\multicolumn{10}{c}{\textit{\textbf{Dense Retrievers}}} \\
	\hline
	Qwen3-Embedding-8B & 53.00 & 37.70 & 13.46 & 76.46 & 72.37 & 4.52 & 73.22 & 61.96 & 12.11 \\
	\rowcolor{gray!8} Qwen3-Embedding-4B & 52.30 & 36.48 & 13.60 & 73.92 & 69.46 & 4.70 & 65.09 & 49.55 & 17.47 \\
	Qwen3-Embedding-0.6B & 53.65 & 34.47 & 16.77 & 72.78 & 64.86 & 8.60 & 53.82 & 35.81 & 20.66 \\
	\rowcolor{gray!8} llama-embed-nemotron-8b & 64.55 & 46.21 & 17.70 & 83.91 & 80.74 & 3.29 & 71.10 & 57.51 & 14.96 \\
	llama-nemotron-embed-1b-v2 & 63.99 & 45.81 & 17.08 & 84.11 & 82.34 & 1.89 & 72.15 & 49.47 & 24.20 \\
	\rowcolor{gray!8} pplx-embed-v1-4b & 46.65 & 33.24 & 9.48 & 51.51 & 49.58 & 1.72 & 25.59 & 18.81 & 6.24 \\
	pplx-embed-v1-0.6b & 58.06 & 40.28 & 15.47 & 72.11 & 68.35 & 3.80 & 43.09 & 30.10 & 12.13 \\
	\rowcolor{gray!8} jina-embeddings-v5-small & 49.55 & 32.90 & 12.48 & 65.47 & 60.79 & 4.77 & 51.63 & 35.65 & 17.91 \\
	jina-embeddings-v5-nano & 48.42 & 31.96 & 12.04 & 67.18 & 62.14 & 5.18 & 53.26 & 35.91 & 19.90 \\
	\rowcolor{gray!8} jina-embeddings-v3 & 42.00 & 15.96 & 17.26 & 36.65 & 30.03 & 5.55 & 24.98 & 12.51 & 18.01 \\
	gte-Qwen2-7B-instruct & 57.98 & 37.73 & 19.36 & 59.16 & 54.92 & 5.12 & 34.72 & 22.82 & 11.66 \\
	\rowcolor{gray!8} gte-Qwen2-1.5B-instruct & 80.78 & 48.41 & 34.51 & 93.08 & 83.40 & 10.04 & 66.11 & 42.92 & 27.53 \\
	gte-multilingual-base & 51.82 & 32.56 & 15.50 & 53.77 & 49.03 & 4.77 & 18.04 & 13.31 & 4.05 \\
	\rowcolor{gray!8} snowflake-arctic-embed-l-v2.0 & 63.79 & 41.13 & 21.71 & 78.62 & 73.05 & 5.86 & 58.52 & 41.86 & 18.42 \\
	snowflake-arctic-embed-m-v2.0 & 71.93 & 41.05 & 31.37 & 86.41 & 75.76 & 10.73 & 68.23 & 41.51 & 28.81 \\
	\rowcolor{gray!8} multilingual-e5-large & 96.15 & 58.94 & 39.27 & 99.92 & 93.54 & 6.40 & 62.76 & 50.32 & 12.42 \\
	multilingual-e5-base & 94.73 & 55.47 & 41.76 & 99.68 & 90.80 & 8.91 & 62.10 & 46.71 & 15.64 \\
	\rowcolor{gray!8} multilingual-e5-small & 95.91 & 53.53 & 44.64 & 99.73 & 90.29 & 9.47 & 44.14 & 33.28 & 8.89 \\
	harrier-oss-v1-0.6b & 95.66 & 58.41 & 39.36 & 99.75 & 92.56 & 7.23 & 81.70 & 60.81 & 22.36 \\
	\rowcolor{gray!8} harrier-oss-v1-270m & 94.17 & 54.10 & 42.66 & 99.56 & 89.94 & 9.69 & 80.99 & 59.33 & 23.30 \\
	voyage-4-nano & 52.86 & 34.77 & 13.71 & 57.84 & 54.21 & 3.29 & 37.41 & 26.55 & 11.22 \\
	\rowcolor{gray!8} embeddinggemma-300m & 67.44 & 43.81 & 23.97 & 86.35 & 79.36 & 7.17 & 53.22 & 35.49 & 19.59 \\
	bge-m3 & 53.79 & 35.95 & 14.19 & 64.42 & 60.49 & 3.90 & 39.25 & 28.71 & 10.07 \\
	\rowcolor{gray!8} nomic-embed-text-v2-moe & 67.59 & 43.82 & 23.79 & 82.01 & 76.42 & 5.69 & 64.86 & 47.45 & 18.82 \\
	granite-embedding-278m & 58.36 & 31.70 & 22.69 & 64.49 & 54.94 & 9.01 & 44.14 & 25.67 & 20.64 \\
	\rowcolor{gray!8} paraphrase-multilingual-MiniLM & 36.88 & 14.47 & 12.62 & 23.47 & 18.55 & 3.80 & 11.63 & 3.42 & 12.00 \\
	LaBSE & 37.53 & 12.04 & 13.36 & 24.03 & 18.08 & 4.12 & 8.96 & 3.58 & 6.16 \\
	\hline
	\multicolumn{10}{c}{\textit{\textbf{Sparse Retrievers}}} \\
	\hline
	opensearch-neural-sparse-v1 & 91.03 & 40.92 & 50.69 & 95.24 & 71.30 & 24.36 & 71.66 & 36.16 & 40.32 \\
	\rowcolor{gray!8} BM25 (XLM-R tokenizer) & 93.68 & 45.08 & 52.07 & 98.75 & 87.18 & 11.98 & 89.76 & 49.96 & 43.72 \\
	bge-m3-sparse & 87.77 & 45.72 & 44.25 & 96.02 & 86.13 & 10.22 & 74.86 & 54.16 & 22.90 \\
	\hline
	\multicolumn{10}{c}{\textit{\textbf{Late-Interaction Retrievers}}} \\
	\hline
	jina-colbert-v2 & 60.42 & 45.96 & 13.49 & 79.10 & 75.43 & 3.65 & 66.93 & 46.97 & 20.17 \\
	\rowcolor{gray!8} LFM2-ColBERT-350M & 90.73 & 55.05 & 38.53 & 98.41 & 80.51 & 18.02 & 87.75 & 41.71 & 49.71 \\
	bge-m3-colbert & 62.02 & 44.96 & 14.93 & 79.75 & 76.97 & 2.75 & 53.07 & 41.40 & 11.46 \\
	\hline
	\end{tabular}
	}
\end{table*}

%%%%%%%%%%%%%%%%%%%%%%%%%%%%%%%%%%%%%%%%%%%%%%%%%%%%%%%%%%%%
\clearpage
\section*{NeurIPS Paper Checklist}

\begin{enumerate}

\item {\bf Claims}
    \item[] Question: Do the main claims made in the abstract and introduction accurately reflect the paper's contributions and scope?
    \item[] Answer: \answerYes{} % Replace by \answerYes{}, \answerNo{}, or \answerNA{}.
    \item[] Justification: Abstract and Section 1 completely reflects the paper's contribution.
    \item[] Guidelines:
    \begin{itemize}
        \item The answer \answerNA{} means that the abstract and introduction do not include the claims made in the paper.
        \item The abstract and/or introduction should clearly state the claims made, including the contributions made in the paper and important assumptions and limitations. A \answerNo{} or \answerNA{} answer to this question will not be perceived well by the reviewers. 
        \item The claims made should match theoretical and experimental results, and reflect how much the results can be expected to generalize to other settings. 
        \item It is fine to include aspirational goals as motivation as long as it is clear that these goals are not attained by the paper. 
    \end{itemize}

\item {\bf Limitations}
    \item[] Question: Does the paper discuss the limitations of the work performed by the authors?
    \item[] Answer: \answerYes{} % Replace by \answerYes{}, \answerNo{}, or \answerNA{}.
    \item[] Justification: The paper discusses the limitations of the work in Appendix A.
    \item[] Guidelines:
    \begin{itemize}
        \item The answer \answerNA{} means that the paper has no limitation while the answer \answerNo{} means that the paper has limitations, but those are not discussed in the paper. 
        \item The authors are encouraged to create a separate ``Limitations'' section in their paper.
        \item The paper should point out any strong assumptions and how robust the results are to violations of these assumptions (e.g., independence assumptions, noiseless settings, model well-specification, asymptotic approximations only holding locally). The authors should reflect on how these assumptions might be violated in practice and what the implications would be.
        \item The authors should reflect on the scope of the claims made, e.g., if the approach was only tested on a few datasets or with a few runs. In general, empirical results often depend on implicit assumptions, which should be articulated.
        \item The authors should reflect on the factors that influence the performance of the approach. For example, a facial recognition algorithm may perform poorly when image resolution is low or images are taken in low lighting. Or a speech-to-text system might not be used reliably to provide closed captions for online lectures because it fails to handle technical jargon.
        \item The authors should discuss the computational efficiency of the proposed algorithms and how they scale with dataset size.
        \item If applicable, the authors should discuss possible limitations of their approach to address problems of privacy and fairness.
        \item While the authors might fear that complete honesty about limitations might be used by reviewers as grounds for rejection, a worse outcome might be that reviewers discover limitations that aren't acknowledged in the paper. The authors should use their best judgment and recognize that individual actions in favor of transparency play an important role in developing norms that preserve the integrity of the community. Reviewers will be specifically instructed to not penalize honesty concerning limitations.
    \end{itemize}

\item {\bf Theory assumptions and proofs}
    \item[] Question: For each theoretical result, does the paper provide the full set of assumptions and a complete (and correct) proof?
    \item[] Answer: \answerNA{} % Replace by \answerYes{}, \answerNo{}, or \answerNA{}.
    \item[] Justification: This paper does not involve theoretical assumptions and proofs.
    \item[] Guidelines:
    \begin{itemize}
        \item The answer \answerNA{} means that the paper does not include theoretical results. 
        \item All the theorems, formulas, and proofs in the paper should be numbered and cross-referenced.
        \item All assumptions should be clearly stated or referenced in the statement of any theorems.
        \item The proofs can either appear in the main paper or the supplemental material, but if they appear in the supplemental material, the authors are encouraged to provide a short proof sketch to provide intuition. 
        \item Inversely, any informal proof provided in the core of the paper should be complemented by formal proofs provided in appendix or supplemental material.
        \item Theorems and Lemmas that the proof relies upon should be properly referenced. 
    \end{itemize}

    \item {\bf Experimental result reproducibility}
    \item[] Question: Does the paper fully disclose all the information needed to reproduce the main experimental results of the paper to the extent that it affects the main claims and/or conclusions of the paper (regardless of whether the code and data are provided or not)?
    \item[] Answer: \answerYes{} % Replace by \answerYes{}, \answerNo{}, or \answerNA{}.
    \item[] Justification: This is disclosed in Appendix C, Appendix D, Appendix E.
    \item[] Guidelines:
    \begin{itemize}
        \item The answer \answerNA{} means that the paper does not include experiments.
        \item If the paper includes experiments, a \answerNo{} answer to this question will not be perceived well by the reviewers: Making the paper reproducible is important, regardless of whether the code and data are provided or not.
        \item If the contribution is a dataset and\slash or model, the authors should describe the steps taken to make their results reproducible or verifiable. 
        \item Depending on the contribution, reproducibility can be accomplished in various ways. For example, if the contribution is a novel architecture, describing the architecture fully might suffice, or if the contribution is a specific model and empirical evaluation, it may be necessary to either make it possible for others to replicate the model with the same dataset, or provide access to the model. In general. releasing code and data is often one good way to accomplish this, but reproducibility can also be provided via detailed instructions for how to replicate the results, access to a hosted model (e.g., in the case of a large language model), releasing of a model checkpoint, or other means that are appropriate to the research performed.
        \item While NeurIPS does not require releasing code, the conference does require all submissions to provide some reasonable avenue for reproducibility, which may depend on the nature of the contribution. For example
        \begin{enumerate}
            \item If the contribution is primarily a new algorithm, the paper should make it clear how to reproduce that algorithm.
            \item If the contribution is primarily a new model architecture, the paper should describe the architecture clearly and fully.
            \item If the contribution is a new model (e.g., a large language model), then there should either be a way to access this model for reproducing the results or a way to reproduce the model (e.g., with an open-source dataset or instructions for how to construct the dataset).
            \item We recognize that reproducibility may be tricky in some cases, in which case authors are welcome to describe the particular way they provide for reproducibility. In the case of closed-source models, it may be that access to the model is limited in some way (e.g., to registered users), but it should be possible for other researchers to have some path to reproducing or verifying the results.
        \end{enumerate}
    \end{itemize}

\item {\bf Open access to data and code}
    \item[] Question: Does the paper provide open access to the data and code, with sufficient instructions to faithfully reproduce the main experimental results, as described in supplemental material?
    \item[] Answer: \answerYes{} % Replace by \answerYes{}, \answerNo{}, or \answerNA{}.
    \item[] Justification: The evaluation code, setup scripts, and exact commands are released at \url{https://github.com/yjoonjang/MLAIRE}, and the datasets are released on Hugging Face at \url{https://huggingface.co/collections/yjoonjang/mlaire} (Appendix~\ref{app:infrastructure}).
    \item[] Guidelines:
    \begin{itemize}
        \item The answer \answerNA{} means that paper does not include experiments requiring code.
        \item Please see the NeurIPS code and data submission guidelines (\url{https://neurips.cc/public/guides/CodeSubmissionPolicy}) for more details.
        \item While we encourage the release of code and data, we understand that this might not be possible, so \answerNo{} is an acceptable answer. Papers cannot be rejected simply for not including code, unless this is central to the contribution (e.g., for a new open-source benchmark).
        \item The instructions should contain the exact command and environment needed to run to reproduce the results. See the NeurIPS code and data submission guidelines (\url{https://neurips.cc/public/guides/CodeSubmissionPolicy}) for more details.
        \item The authors should provide instructions on data access and preparation, including how to access the raw data, preprocessed data, intermediate data, and generated data, etc.
        \item The authors should provide scripts to reproduce all experimental results for the new proposed method and baselines. If only a subset of experiments are reproducible, they should state which ones are omitted from the script and why.
        \item At submission time, to preserve anonymity, the authors should release anonymized versions (if applicable).
        \item Providing as much information as possible in supplemental material (appended to the paper) is recommended, but including URLs to data and code is permitted.
    \end{itemize}

\item {\bf Experimental setting/details}
    \item[] Question: Does the paper specify all the training and test details (e.g., data splits, hyperparameters, how they were chosen, type of optimizer) necessary to understand the results?
    \item[] Answer: \answerYes{} % Replace by \answerYes{}, \answerNo{}, or \answerNA{}.
    \item[] Justification: This is specified in Section 4 and Appendix C, Appendix D, Appendix E.
    \item[] Guidelines:
    \begin{itemize}
        \item The answer \answerNA{} means that the paper does not include experiments.
        \item The experimental setting should be presented in the core of the paper to a level of detail that is necessary to appreciate the results and make sense of them.
        \item The full details can be provided either with the code, in appendix, or as supplemental material.
    \end{itemize}

\item {\bf Experiment statistical significance}
    \item[] Question: Does the paper report error bars suitably and correctly defined or other appropriate information about the statistical significance of the experiments?
    \item[] Answer: \answerYes{} % Replace by \answerYes{}, \answerNo{}, or \answerNA{}.
    \item[] Justification: Appendix~\ref{app:significance} reports exact Pearson and Spearman tests with Fisher-$z$ 95\% confidence intervals for the association between nDCG and LPR, and Wilson 95\% confidence intervals for per-model LPR, together with a conservative bound that accounts for queries sharing a content group. The evaluated retrievers are fixed public checkpoints under a deterministic protocol, so we do not report variability over training seeds.
    \item[] Guidelines:
    \begin{itemize}
        \item The answer \answerNA{} means that the paper does not include experiments.
        \item The authors should answer \answerYes{} if the results are accompanied by error bars, confidence intervals, or statistical significance tests, at least for the experiments that support the main claims of the paper.
        \item The factors of variability that the error bars are capturing should be clearly stated (for example, train/test split, initialization, random drawing of some parameter, or overall run with given experimental conditions).
        \item The method for calculating the error bars should be explained (closed form formula, call to a library function, bootstrap, etc.)
        \item The assumptions made should be given (e.g., Normally distributed errors).
        \item It should be clear whether the error bar is the standard deviation or the standard error of the mean.
        \item It is OK to report 1-sigma error bars, but one should state it. The authors should preferably report a 2-sigma error bar than state that they have a 96\% CI, if the hypothesis of Normality of errors is not verified.
        \item For asymmetric distributions, the authors should be careful not to show in tables or figures symmetric error bars that would yield results that are out of range (e.g., negative error rates).
        \item If error bars are reported in tables or plots, the authors should explain in the text how they were calculated and reference the corresponding figures or tables in the text.
    \end{itemize}

\item {\bf Experiments compute resources}
    \item[] Question: For each experiment, does the paper provide sufficient information on the computer resources (type of compute workers, memory, time of execution) needed to reproduce the experiments?
    \item[] Answer: \answerYes{} % Replace by \answerYes{}, \answerNo{}, or \answerNA{}.
    \item[] Justification: We detail the hardware specifications, including the compute workers (GPUs) and system memory used for all evaluations, in Appendix D.
    \item[] Guidelines:
    \begin{itemize}
        \item The answer \answerNA{} means that the paper does not include experiments.
        \item The paper should indicate the type of compute workers CPU or GPU, internal cluster, or cloud provider, including relevant memory and storage.
        \item The paper should provide the amount of compute required for each of the individual experimental runs as well as estimate the total compute. 
        \item The paper should disclose whether the full research project required more compute than the experiments reported in the paper (e.g., preliminary or failed experiments that didn't make it into the paper). 
    \end{itemize}
    
\item {\bf Code of ethics}
    \item[] Question: Does the research conducted in the paper conform, in every respect, with the NeurIPS Code of Ethics \url{https://neurips.cc/public/EthicsGuidelines}?
    \item[] Answer: \answerYes{} % Replace by \answerYes{}, \answerNo{}, or \answerNA{}.
    \item[] Justification: Our research strictly conforms to the Code of Ethics. We solely utilize publicly available datasets and provide a comprehensive discussion of dataset licenses, fairness trade-offs, and the responsible use of our proposed metrics in Appendix B.
    \item[] Guidelines:
    \begin{itemize}
        \item The answer \answerNA{} means that the authors have not reviewed the NeurIPS Code of Ethics.
        \item If the authors answer \answerNo, they should explain the special circumstances that require a deviation from the Code of Ethics.
        \item The authors should make sure to preserve anonymity (e.g., if there is a special consideration due to laws or regulations in their jurisdiction).
    \end{itemize}

\item {\bf Broader impacts}
    \item[] Question: Does the paper discuss both potential positive societal impacts and negative societal impacts of the work performed?
    \item[] Answer: \answerYes{} % Replace by \answerYes{}, \answerNo{}, or \answerNA{}.
    \item[] Justification: We explicitly address both the potential positive impacts and the negative societal implications, including potential harms and mitigation strategies for responsible deployment, in Appendix B (Ethical Considerations).
    \item[] Guidelines:
    \begin{itemize}
        \item The answer \answerNA{} means that there is no societal impact of the work performed.
        \item If the authors answer \answerNA{} or \answerNo, they should explain why their work has no societal impact or why the paper does not address societal impact.
        \item Examples of negative societal impacts include potential malicious or unintended uses (e.g., disinformation, generating fake profiles, surveillance), fairness considerations (e.g., deployment of technologies that could make decisions that unfairly impact specific groups), privacy considerations, and security considerations.
        \item The conference expects that many papers will be foundational research and not tied to particular applications, let alone deployments. However, if there is a direct path to any negative applications, the authors should point it out. For example, it is legitimate to point out that an improvement in the quality of generative models could be used to generate Deepfakes for disinformation. On the other hand, it is not needed to point out that a generic algorithm for optimizing neural networks could enable people to train models that generate Deepfakes faster.
        \item The authors should consider possible harms that could arise when the technology is being used as intended and functioning correctly, harms that could arise when the technology is being used as intended but gives incorrect results, and harms following from (intentional or unintentional) misuse of the technology.
        \item If there are negative societal impacts, the authors could also discuss possible mitigation strategies (e.g., gated release of models, providing defenses in addition to attacks, mechanisms for monitoring misuse, mechanisms to monitor how a system learns from feedback over time, improving the efficiency and accessibility of ML).
    \end{itemize}
    
\item {\bf Safeguards}
    \item[] Question: Does the paper describe safeguards that have been put in place for responsible release of data or models that have a high risk for misuse (e.g., pre-trained language models, image generators, or scraped datasets)?
    \item[] Answer: \answerNA{} % Replace by \answerYes{}, \answerNo{}, or \answerNA{}.
    \item[] Justification: We answered \answerNA{} as this work proposes an evaluation benchmark derived from established public datasets, which poses no dual-use risks or high potential for misuse.
    \item[] Guidelines:
    \begin{itemize}
        \item The answer \answerNA{} means that the paper poses no such risks.
        \item Released models that have a high risk for misuse or dual-use should be released with necessary safeguards to allow for controlled use of the model, for example by requiring that users adhere to usage guidelines or restrictions to access the model or implementing safety filters. 
        \item Datasets that have been scraped from the Internet could pose safety risks. The authors should describe how they avoided releasing unsafe images.
        \item We recognize that providing effective safeguards is challenging, and many papers do not require this, but we encourage authors to take this into account and make a best faith effort.
    \end{itemize}

\item {\bf Licenses for existing assets}
    \item[] Question: Are the creators or original owners of assets (e.g., code, data, models), used in the paper, properly credited and are the license and terms of use explicitly mentioned and properly respected?
    \item[] Answer: \answerYes{} % Replace by \answerYes{}, \answerNo{}, or \answerNA{}.
    \item[] Justification: We cite the original papers and model releases for all datasets and retriever checkpoints used in the evaluation. Dataset licenses are detailed in Appendix B. All evaluated retrievers are publicly available checkpoints used according to their respective model cards and licenses. The released MLAIRE assets follow the reuse requirements of the underlying datasets.
    \item[] Guidelines:
    \begin{itemize}
        \item The answer \answerNA{} means that the paper does not use existing assets.
        \item The authors should cite the original paper that produced the code package or dataset.
        \item The authors should state which version of the asset is used and, if possible, include a URL.
        \item The name of the license (e.g., CC-BY 4.0) should be included for each asset.
        \item For scraped data from a particular source (e.g., website), the copyright and terms of service of that source should be provided.
        \item If assets are released, the license, copyright information, and terms of use in the package should be provided. For popular datasets, \url{paperswithcode.com/datasets} has curated licenses for some datasets. Their licensing guide can help determine the license of a dataset.
        \item For existing datasets that are re-packaged, both the original license and the license of the derived asset (if it has changed) should be provided.
        \item If this information is not available online, the authors are encouraged to reach out to the asset's creators.
    \end{itemize}

\item {\bf New assets}
    \item[] Question: Are new assets introduced in the paper well documented and is the documentation provided alongside the assets?
    \item[] Answer: \answerYes{} % Replace by \answerYes{}, \answerNo{}, or \answerNA{}.
    \item[] Justification: The newly constructed MLAIRE benchmark and the evaluation code are thoroughly documented in the main text and appendices. They are publicly released at \url{https://huggingface.co/collections/yjoonjang/mlaire} (datasets, with dataset cards and Croissant metadata) and \url{https://github.com/yjoonjang/MLAIRE} (code).
    \item[] Guidelines:
    \begin{itemize}
        \item The answer \answerNA{} means that the paper does not release new assets.
        \item Researchers should communicate the details of the dataset\slash code\slash model as part of their submissions via structured templates. This includes details about training, license, limitations, etc. 
        \item The paper should discuss whether and how consent was obtained from people whose asset is used.
        \item At submission time, remember to anonymize your assets (if applicable). You can either create an anonymized URL or include an anonymized zip file.
    \end{itemize}

\item {\bf Crowdsourcing and research with human subjects}
    \item[] Question: For crowdsourcing experiments and research with human subjects, does the paper include the full text of instructions given to participants and screenshots, if applicable, as well as details about compensation (if any)? 
    \item[] Answer: \answerNA{} % Replace by \answerYes{}, \answerNo{}, or \answerNA{}.
    \item[] Justification: We answered \answerNA{} because this research relies entirely on automated evaluations using existing benchmark datasets (Belebele, XQuAD, MLQA) and does not involve any crowdsourcing or human subject experiments.
    \item[] Guidelines:
    \begin{itemize}
        \item The answer \answerNA{} means that the paper does not involve crowdsourcing nor research with human subjects.
        \item Including this information in the supplemental material is fine, but if the main contribution of the paper involves human subjects, then as much detail as possible should be included in the main paper. 
        \item According to the NeurIPS Code of Ethics, workers involved in data collection, curation, or other labor should be paid at least the minimum wage in the country of the data collector. 
    \end{itemize}

\item {\bf Institutional review board (IRB) approvals or equivalent for research with human subjects}
    \item[] Question: Does the paper describe potential risks incurred by study participants, whether such risks were disclosed to the subjects, and whether Institutional Review Board (IRB) approvals (or an equivalent approval/review based on the requirements of your country or institution) were obtained?
    \item[] Answer: \answerNA{}
    \item[] Justification: We answered \answerNA{} because this study does not involve human subjects or participants, making IRB approval inapplicable.
    \item[] Guidelines:
    \begin{itemize}
        \item The answer \answerNA{} means that the paper does not involve crowdsourcing nor research with human subjects.
        \item Depending on the country in which research is conducted, IRB approval (or equivalent) may be required for any human subjects research. If you obtained IRB approval, you should clearly state this in the paper. 
        \item We recognize that the procedures for this may vary significantly between institutions and locations, and we expect authors to adhere to the NeurIPS Code of Ethics and the guidelines for their institution. 
        \item For initial submissions, do not include any information that would break anonymity (if applicable), such as the institution conducting the review.
    \end{itemize}

\item {\bf Declaration of LLM usage}
    \item[] Question: Does the paper describe the usage of LLMs if it is an important, original, or non-standard component of the core methods in this research? Note that if the LLM is used only for writing, editing, or formatting purposes and does \emph{not} impact the core methodology, scientific rigor, or originality of the research, declaration is not required.
    %this research? 
    \item[] Answer: \answerYes{} % Replace by \answerYes{}, \answerNo{}, or \answerNA{}.
    \item[] Justification: We use Qwen2.5-7B-Instruct only as the generator in the controlled RAG motivation experiment described in Section 2.2. LLMs are not used to construct MLAIRE, define the evaluation labels, or implement the proposed retrieval metrics. Any LLM assistance for writing or formatting does not affect the core methodology.
    \item[] Guidelines:
    \begin{itemize}
        \item The answer \answerNA{} means that the core method development in this research does not involve LLMs as any important, original, or non-standard components.
        \item Please refer to our LLM policy in the NeurIPS handbook for what should or should not be described.
    \end{itemize}

\end{enumerate}

\end{document}